\documentclass[12pt]{article}
\usepackage{amsmath}
\usepackage{graphicx,psfrag,epsf}
\usepackage{enumerate}
\usepackage{natbib}
\usepackage{bm}
\usepackage{hyperref}
\usepackage{url} 
\usepackage{amssymb}
\newcommand{\blind}{0}

\begin{document}

\def\spacingset#1{\renewcommand{\baselinestretch}%
{#1}\small\normalsize} \spacingset{1}


\if0\blind
{
  \title{\bf Out of Sample Predictability in Predictive Regressions with Many Predictor Candidates}
  \author{	Jes\'{u}s Gonzalo\thanks{
    The authors gratefully acknowledge financial support 
    from the {\it Spanish Ministerio de Economica y Competitividad} through 
    grants PID2019-104960GB-I00, TED2021-129784B-I00 and the UK {\it Economic and Social Research Council} through grant ES/W000989/1}
\hspace{.2cm}\\
    Department of Economics, Universidad Carlos III de Madrid\\
    and \\
    Jean-Yves Pitarakis \\
    Department of Economics, University of Southampton}
  \maketitle
} \fi

\if1\blind
{
  \bigskip
  \bigskip
  \bigskip
  \begin{center}
    {\LARGE\bf Title}
\end{center}
  \medskip
} \fi

\bigskip
\begin{abstract}
This paper is concerned with detecting the presence of out of sample predictability in linear predictive regressions with 
a potentially large set of candidate predictors. We propose a procedure based on out of sample MSE comparisons that is 
implemented in a pairwise manner using one predictor at a time and resulting in an aggregate test statistic that is standard normally distributed under the global null hypothesis of no linear predictability. Predictors can be highly persistent, purely stationary or a 
combination of both. Upon rejection of the null hypothesis we subsequently introduce a predictor screening procedure designed to identify the most active predictors. An empirical application 
to key predictors of US economic activity illustrates the usefulness of our methods and highlights the
important forward looking role played by the series of manufacturing new orders. 
\end{abstract}

\noindent%
\vfill

\newpage
\spacingset{1.8} 
\section{Introduction}
\label{sec:intro}

Comparing the out of sample predictive accuracy of competing statistical models in data rich environments is an essential component of data science and a key step in the workflow that aims to produce reliable forecasts of an outcome of interest or discriminate between competing hypotheses. A vast body of statistics research over the past decade has been concerned with developing estimation and prediction techniques that can accommodate the availability of large datasets via regularisation techniques and sparsity assumptions on the underlying DGPs. An important objective driving this literature has been to obtain accurate out of sample predictions of a response variable via suitable estimation and covariate selection techniques. 

The detection of predictability within linear regression settings has also been the subject of extensive research in the traditional econometrics literature. The broadly labelled topic of predictive regressions for instance has become an important field of research in its own right due to the specificities associated with economic data and the complications that these may cause for estimation, inference and prediction (e.g. persistent nature of many financial and economic predictors, endogeneity considerations, low signal to noise ratios, imbalance in the persistence properties of predictand and predictors). Unlike the above mentioned statistics literature however, predictive regressions as explored in econometrics have been mainly concerned with in-sample goodness of fit measures and traditional significance testing designed to explicitly accommodate these specificities, often in the context of single predictor settings (see Gonzalo and Pitarakis (2019) for a survey of this literature). Some of the early applied research did also highlight the importance of distinguishing between in-sample and out-of-sample predictability in the context of stock return predictability with valuation ratios and related predictors (see Pesaran and Timmermann (1995), Goyal and Welch (2008)).  

Our objective in this paper is to consider this predictive regression environment as commonly explored in the econometrics literature and propose a method for detecting the potential presence of out-of-sample linear predictability when the latter is induced by one or more predictors from a potentially large pool of candidate predictors, possibly exceeding the available sample size. These predictors could be purely stationary or highly persistent without affecting the validity of our proposed approach and without the need for the investigator to have knowledge of these properties. 
We are thinking of an environment  where one is confronted with not only a
potentially large pool of predictors but also with these predictors allowed to display a mixture of dynamic
characteristics, some (or all) being highly persistent and others noisier as it commonly occurs in economic and financial data. A macroeconomist interested in forecasts of GDP growth for instance faces hundreds of
potentially useful predictors ranging from noisy indicators with very little memory such as financial returns
to more persistent series such as interest rates. With an increased availability of unconventional predictor candidates that go beyond traditional macro and financial series (e.g., internet search related data, climate related predictors,  etc.)
and whose persistence properties are not well known,  this aspect of our proposed methods that allows one to remain agnostic about the stochastic properties of predictors is particularly important. 

Our operating environment is that of a potentially large number of nested specifications that also include an intercept only model which we view as the benchmark model or the maintained theory. More specifically, we focus our attention on testing this benchmark specification or global null against the alternative hypothesis that at least one of the predictors under consideration is active in the sense of improving out of sample MSEs relative to the benchmark. 

The approach introduced in this paper is able to accommodate a large number of predictors as it relies on 
multiple pairwise comparisons of the benchmark model with a larger model that includes solely one predictor at a time. These pairwise MSE comparisons are implemented via the repeated evaluation of a novel test statistic suitable for out of sample predictive accuracy comparisons in nested environments. This latter aspect is a particularly important contribution of this research as traditional test statistics for predictive accuracy comparisons are known to fail in nested 
environments. This is due to a variance degeneracy problem induced by the fact that under the null hypothesis the population forecast errors of the two models being compared are asymptotically identical. Our proposed test statistic is designed in such a way that nestedness induced variance degeneracy is bypassed. Each pairwise MSE comparison between the benchmark model (global null) and a specification that includes one predictor at a time is performed via this novel test statistic. The resulting 
individual test statistics (as many as there are predictors within the aggregate pool of predictors) are subsequently reassembled into a single aggregate statistic allowing us to test the global null of no predictability against the 
alternative that at least one of the predictors is active. 

Upon rejection of the benchmark model the important question as to which predictors are the most important drivers of predictability also arises. To address this question we subsequently introduce a covariate screening method that allows us to identify the key predictor that most improves the accuracy of 
forecasts of the response variable relative to the forecasts based on the benchmark model. We refer to such a predictor as the key player. Although the identification of a single predictor may 
come across as providing only a limited picture of an underlying {\it true} specification, it is nevertheless a valuable picture. Parsimonious models can often achieve desirable levels of predictive strength by avoiding distortions due to overfitting for instance.

Our operating environment is particularly relevant to economic and financial applications where one is interested in the maintained hypothesis of no predictability whereby the response variable of interest is best described by a martingale difference process (e.g. excess stock returns, currency returns, consumption growth). More generally, one is often faced with the need to compare the predictive accuracy of a simple model nested within a richer one. Nested models are one of
the most commonly encountered setting in empirical research and help answer fundamental questions such
as: does the inclusion of additional predictors significantly improve the predictive power of a smaller
model or a non-predictability benchmark? 

The plan of the paper is as follows. Section 2 introduces our modelling environment and the key test statistics that are used to implement our predictive accuracy comparisons. Section 3 develops the asymptotic theory under the global null and is followed by a comprehensive local power analysis in Section 4. Section 5 introduces a theoretically supported power enhancing transformation to our proposed tests. Section 6 introduces our key player estimator and studies its asymptotic properties.  Section 7 demonstrates the finite sample properties of our methods through a comprehensive simulation based exercise. Finally Section 8 illustrates the usefulness of our methods through an application to the predictability of US economic activity. Section 9 concludes. All technical proofs and further simulations are placed in a supplementary appendix. 

\vspace{-0.5cm} 

\section{Models and Theory}

Let $\{y_{t}\}$ denote a scalar random process. Given a sample of size $n$ we wish to assess the presence of linear one-step ahead predictability in $y_{t}$. If present, predictability is  induced by at least one predictor from a finite pool of $p$ predictors ${\bm x}_{t}=(x_{1t},\ldots,x_{pt})'$. 
Predictability is understood to be present whenever an intercept only benchmark model (the global null) is rejected in favour of a larger model on the basis of out-of-sample MSE based comparisons. Thus the generic 
framework within which we operate is given by the predictive regressions \vspace{-0.2cm}
\begin{eqnarray}
	y_{t+1} & = & \theta_{0} + {\bm \beta}'{\bm x}_{t}+u_{t+1} \label{eq:eq1}
\end{eqnarray}
where ${\bm \beta}=(\beta_{1},\ldots,\beta_{p})'$ and $u_{t}$ is a 
random disturbance term. For later use we also define ${\bm \theta}=(\theta_{0},{\bm \beta}')'$ and 
${\bm w}_{t}=(1,{\bm x}_{t}')'$ so that (\ref{eq:eq1}) can equivalently be expressed as 
$y_{t+1}={\bm \theta}'{\bm w}_{t}+u_{t+1}$. The predictors collected in ${\bm x}_{t}$ may have different degrees of persistence as commonly encountered in economic applications. Nevertheless our approach does not rely on any knowledge of the persistence properties of the pool of predictors available to the investigator.
We view (\ref{eq:eq1}) as encompassing a family of nested linear predictive regressions including the benchmark specification given by \vspace{-0.4cm}
\begin{eqnarray}
	y_{t+1} & = & \theta_{0} + u_{t+1}. \label{eq:eq2}
\end{eqnarray}

Given the above framework our main goal is to address the following questions. Suppose a researcher has access to a pool of predictors collected within ${\bm x}_{t}$. Is at least one of these 
predictors active relative to the benchmark model in (\ref{eq:eq2})? In the affirmative, is it possible to accurately identify 
which one of the $p$ predictors has the strongest influence in the sense of improving forecast accuracy the most relative to the benchmark?

To formalise our environment we let $\hat{y}_{0,t+1|t}$ denote the one-step ahead forecasts of $y_{t+1}$ obtained from the 
benchmark model in (\ref{eq:eq2}) and $\hat{y}_{j,t+1|t}$, $j=1,\ldots,p$, the one-step ahead forecasts of $y_{t+1}$ obtained from (\ref{eq:eq1}) using
one predictor at a time from the available collection of $p$ predictors and inclusive of a fitted intercept. The corresponding forecast errors are $\hat{e}_{0,t+1|t}=y_{t+1}-\hat{y}_{0,t+1|t}$ and $\hat{e}_{j,t+1|t}=y_{t+1}-\hat{y}_{j,t+1|t}$. Out of sample 
forecasts are constructed recursively with an expanding window approach. We estimate each predictive regression via recursive least-squares starting from an initial window of size $t=1,\ldots,k_{0}$ and progressively expanding the estimation window up to $n-1$. Throughout this paper $k_{0}$ is taken to be a given a fraction $\pi_{0}$ of the sample size and we write $k_{0}=[n\pi_{0}]$ for some $\pi_{0}\in (0,1)$. Under the benchmark model we have 
$\hat{\theta}_{0t}=\sum_{s=1}^{t}y_{s}/t$ leading to the unconditional mean forecasts $\hat{y}_{0,t+1|t}=\hat{\theta}_{0t}$. Under the larger models estimated with an intercept and one predictor at a time 
we have $\hat{\bm \theta}_{jt}=(\sum_{s=1}^{t}\widetilde{\bm w}_{j,s-1}\widetilde{\bm w}_{j,s-1}')^{-1}\sum_{s=1}^{t}\widetilde{\bm w}_{j,s-1}y_{s}$ for $\widetilde{\bm w}_{jt}=(1,w_{jt})$ and $w_{jt} \in \{x_{1t},\ldots,x_{pt}\}$ with forecasts obtained as $\hat{y}_{j,t+1|t}=\hat{\bm \theta}_{jt}' \widetilde{w}_{jt}$ for $t=k_{0},\ldots,n-1$. 
At the end of this pseudo out-of-sample exercise we obtain the $p+1$ sequences of forecast errors $\{\hat{e}_{0,t+1|t}\}_{t=k_{0}}^{n-1}$ and $\{\hat{e}_{j,t+1|t}\}_{t=k_{0}}^{n-1}$ for $j=1,\ldots,p$ which form the basis of our inferences. 
Throughout this paper the maintained null hypothesis is that the population MSEs of the benchmark model and the larger models are equal in the sense that ${\bm \beta}=0$ in (\ref{eq:eq1}) or equivalently model (\ref{eq:eq2}) holds. The alternative of interest is that there is at least one active predictor $w_{jt}$ in the sense that $E[\hat{e}_{0,t+1|t}^{2}-\hat{e}_{j,t+1|t}^{2}]>0$ for at least one $j \in \{1,2,\ldots,p\}$.

Addressing the two questions stated above raises four key challenges which the methods developed in this paper address. The first one arises from the fact that we wish to conduct out of sample 
predictive accuracy comparisons in a nested setting (e.g. intercept only model versus single predictor specifications), rendering traditional sample MSE comparisons ineffective as under the null hypothesis of equal predictive accuracy all forecast errors under consideration will be asymptotically identical, leading to normalised sample 
MSE spreads identically equal to zero in the limit (and similarly for their variances). The second challenge is a dimensionality related complication as we wish our methods to be computationally feasible to implement despite the availability of a potentially large pool of predictors. The third challenge is related to the need for inferences to remain reliable regardless of the persistence properties of the predictors. The fourth challenge has to do with the identification of active predictors upon rejection of the null hypothesis. Although numerous covariate screening procedures have been developed in the statistics literature the validity of most of these relies on assumptions that are not tenable in our time series environment with persistent predictors. 

The issue of predictive performance testing in nested environments has attracted considerable attention in the forecasting literature following the observation that Diebold-Mariano (DM) type constructions (Diebold and Mariano (1995), West (1996)) are not suitable since under the null hypothesis of equal predictive ability the pair of models being compared become identical in the limit. 
Consequently suitably normalised sample MSE spreads and their variance both converge to zero asymptotically resulting in statistics with ill-defined limits. In the context of predictive regressions this problem has been addressed through the use of alternative normalisations of sample MSEs, resulting in test statistics with well defined but non-standard limits requiring bootstrap based approaches (see McCracken (2007), West (2006), Clark and McCracken (2013)). 

More recently, alternative solutions involving modifications to DM type statistics that result in 
conventional standard normal asymptotics regardless of the nested nature of competing models have been developed in Pitarakis (2023). These are similar in spirit to the way Vuong type model selection tests (Vuong (1989)) have been recently adapted to accommodate both nested and non-nested environments via sample splitting and related approaches (see Schennah (2017), Shi (2015) and more recently, Corradi, Fosten and Gutknecht (2023) who focused on predictive accuracy comparisons across overlapping models). The test statistic we introduce in this paper differs from the above and offers yet another useful way of making predictive accuracy comparisons across nested specifications. This novel test statistic is introduced in order to conduct initial pairwise comparisons between the benchmark model in (\ref{eq:eq2}) and the $p$ larger models containing one predictor at a time. The formulation of our test statistics designed to compare the null model with specifications that include one predictor at a time relies on the same 
principles as the statistics introduced in Pitarakis (2023) but does not involve any discarding of sample information. These individual test statistics associated with the $p$ pairwise model comparisons 
are subsequently reassembled into an aggregate statistic designed to test whether at least one of the $p$ predictors is active in predicting $y_{t+1}$. 

The idea of considering one predictor at a time makes the practical implementation of our approach trivial regardless of the 
size of the pool of predictor candidates and is here justified by the fact that our null hypothesis is given by the benchmark model in (\ref{eq:eq2}). This is very much in the spirit of Ghysels et al. (2020) where the authors developed a procedure for testing the statistical significance of a large number of predictors through functionals (e.g. maximum) of multiple individual t-statistics obtained 
from models estimated with one regressor at a time and a benchmark model with none of the explanatory variables included. An important advantage unique to our own setting however is the fact that each of the pairwise DM type statistics will have 
identical limits under our null hypothesis since the two error processes 
across each pairwise comparison will be identical in the limit. This makes the exercise of constructing an aggregate statistic trivial. The average of these individual statistics will by construction also have a distribution that is identical 
to each of its components' distribution. Differently put we do not need to 
be concerned with the behaviour of the covariances of the $p$ individual test statistics and the nested nature of our setting is here used to our advantage. 

Before proceeding further it is also useful to mention the recent but already extensive literature on screening for relevant predictors in high dimensional settings and which is related to our second concern of identifying dominant predictors upon rejection of the benchmark model in (\ref{eq:eq2}). In this context, a particularly popular approach has been based on ranking marginal correlations via marginal linear regressions (see Fan and Lv (2008), McKeague and Qian (2015)). In McKeague and Qian (2015) for instance the authors developed a test for the presence of at least one significant predictor via a maximum correlation type of approach between each predictor and predictand, also assuming a finite pool of predictors. Within our own context and upon rejection of the benchmark model we aim to identify at least one of the active predictors among the pool of $p$ predictors using the above mentioned aggregate test statistic instead. An important aspect accommodated by our framework is the possibility that 
the pool of predictors contains dependent series with different persistence properties in addition to being possibly correlated as it is the norm with economic data.  

We now introduce and motivate the DM type test statistic which will be used to conduct pairwise predictive performance comparisons between the benchmark model and each of the $p$ marginal predictive regression. Recalling that the key complication arising from the underlying nestedness of models is that in the limit $\hat{e}_{0,t+1|t}^{2}$ and $\hat{e}_{j,t+1|t}^{2}$ will be identical under the null hypothesis, we propose to use a sample split estimator for the MSE of the benchmark model instead of the traditional sample mean. This is achieved by splitting the evaluation of the forecast errors associated with the benchmark model 
across two subsamples of size $m_{0}$ and $(n-k_{0})-m_{0}$ respectively. We formulate our test statistic as 
\begin{eqnarray}
	{\cal D}_{n}(m_{0},j) & = & \dfrac{\sqrt{n-k_{0}}}{\hat{\omega}_{n}} \left(\dfrac{1}{2}\left( \dfrac{\sum_{t=k_{0}}^{k_{0}+m_{0}-1}\hat{e}_{0,t+1}^{2}}{m_{0}}+\dfrac{\sum_{t=k_{0}+m_{0}}^{n-1}\hat{e}_{0,t+1}^{2}}{n-k_{0}-m_{0}}\right)- \dfrac{\sum_{t=k_{0}}^{n-1}\hat{e}_{j,t+1}^{2}}{n-k_{0}}\right). \label{eq:eq3}
\end{eqnarray}
 
A simple way to interpret (\ref{eq:eq3}) is by observing that it is based on the difference between the out of sample MSEs of the benchmark and augmented models as it is the case for most predictive accuracy testing statistics. The key novelty here is that we estimate the out of sample MSE of the benchmark model via a sample-split estimator rather than the full sample mean used for the ${\hat{e}_{j,t}}^{2}$'s. In generic notation 
we are essentially estimating an unknown population mean with  
$\overline{X}_{split,n}=(\overline{X}_{1n}+\overline{X}_{2n}))/2$ rather than $\overline{X}_{n}$ and within (\ref{eq:eq3}) $m_{0}$ refers to the chosen location of the sample split. Throughout this paper we take $m_{0}$ to be a user-defined parameter and express it as a fraction $\mu_{0}$ of the effective sample size $n-k_{0}$, writing $m_{0}=[(n-k_{0})\mu_{0}]$. Note of course 
that $\overline{X}_{split,n}$ is identical to $\overline{X}_{n}$ solely if the split occurs in the middle of the sample. A scenario which we will rule out by assumption. 

The motivation for proceeding this way is 
that the formulation in (\ref{eq:eq3}) avoids the variance degeneracy problems associated with nested model comparisons based on traditional DM type formulations. The basic intuition behind the usefulness of (\ref{eq:eq3}) is that 
even if the MSEs of the benchmark and alternative models are identical in the limit, the variances of $\overline{X}_{split,n}$ and $\overline{X}_{n}$ differ thus avoiding the degeneracy problem. 
More specifically, under the null hypothesis the numerator of $(\ref{eq:eq3})$ will have a non-degenerate positive limiting variance provided that $\mu_{0} \in (0,1) \setminus \{1/2\}$. As pointed out above, the exclusion of the case $\mu_{0}=1/2$ is due to the fact that for such a choice of the splitting location we would have $\overline{X}_{split,n}\equiv \overline{X}_{n}$ bringing us back to the traditional DM type constructions which are not suitable for nested model comparisons. We may also think of (\ref{eq:eq3}) as a way of robustifying predictive accuracy inferences to the nestedness/non-nestedness dichotomy. The idea of using averages of subsample means instead of grand means has been used in a variety of other contexts such as the construction of more accurate confidence intervals as discussed in Decrouez and Hall (2014) for instance. 

The normaliser $\hat{\omega}_{n}$ in the denominator of (\ref{eq:eq3}) is understood to be a consistent estimator of the long run variance of the numerator.
Strictly speaking, we may have wished to index it as $\hat{\omega}_{j,n}$ to highlight the fact that it may be estimated using the residuals under the null or the residuals of the single predictor based augmented models (for each $j$). This nuance is naturally inconsequential when it comes to the asymptotics of ${\cal D}_{n}(m_{0},j)$ under the null hypothesis whereby the benchmark model in (\ref{eq:eq2}) holds. It may however have important implications when it comes to the finite sample power properties of our proposed test, an issue we study in greater depth further below.  

Given the sequences $\{\hat{e}_{0,t+1|t}\}_{t=k_{0}}^{n-1}$, $\{\hat{e}_{j,t+1|t}\}_{t=k_{0}}^{n-1}$ and suitable choices for $\mu_{0}$ and $\hat{\omega}_{n}$ the quantities in (\ref{eq:eq3}) can be trivially obtained for each possible predictor $j$, resulting in $p$ such statistics which we aggregate into the following overall statistic \vspace{-0.1cm}
\begin{eqnarray}
	\overline{{\cal D}}_{n}(m_{0}) & = & \dfrac{1}{p} \sum_{j=1}^{p} {\cal D}_{n}(m_{0},j). \label{eq:eq4}
\end{eqnarray}

\noindent A large positive magnitude of $\overline{{\cal D}}_{n}(m_{0})$ is expected to indicate that at least one of the $p$ predictors improves the predictability of $y_{t+1}$ relative to the benchmark model. 

\noindent
{REMARK 1}: One may be tempted to view inferences based on the ${\cal D}_{n}(m_{0},j)$'s through the lens of multiple comparison and {\it reality check} type settings 
and argue for alternative constructs to (\ref{eq:eq4}) such as taking the supremum of these ${\cal D}_{n}(m_{0},j)$'s due for instance to power considerations. It is here important to note however that
within our setting and under the null hypothesis of interest to us all ${\cal D}_{n}(m_{0},j)$'s will be identical in the limit and in turn identical to $\overline{{\cal D}}_{n}(m_{0})$. 
A useful analogy that illustrates this latter point is by thinking of the $p$ components 
of $\overline{{\cal D}}_{n}(m_{0})$ under the {\it global} null hypothesis and in the limit as realisations from the same seed of a random number generator. If we were solely interested in null asymptotics and had no power concerns then any of the ${\cal D}_{n}(m_{0},j)'s$ could be viewed as inter-changeable. 

In what follows our first objective is to establish the limiting behaviour of (\ref{eq:eq4}) under the null hypothesis that there are no active predictors and which we refer to as the global null. We subsequently assess its local power properties against departures from (\ref{eq:eq2}) that are relevant to practitioners. This also allows us to formalise suitable choices for $\mu_{0}$ in the practical implementation of (\ref{eq:eq4}). 

Upon rejection of the null hypothesis the interesting question as to which predictor is the key driver of predictability arises. Although our goal here is not to develop a new covariate screening method, our framework does allow us to identify a key predictor through the analysis of the ${{\cal D}}_{n}(m_{0},j)$ components that make up the test statistic in (\ref{eq:eq4}). We focus our attention on the following estimator \vspace{-0.28cm}
\begin{eqnarray}
	\hat{j}_{n} \in \arg \max_{j=1,\ldots,p} {\cal D}_{n}(m_{0},j) \label{eq:eq5}
\end{eqnarray}
\noindent 
which we expect to be informative about the most important contributor to predictability i.e. the predictor that leads to the greatest reduction in out of sample MSEs relative to the benchmark model and which we refer to as the {\it key player}. A limitation of $\hat{j}_{n}$ is of course
the fact that it allows us to identify only a single predictor. Nevertheless, in numerous economic applications this information 
can be extremely valuable as it isolates the key player that causes the rejection of a maintained martingale difference hypothesis for instance.  

\vspace{-0.4cm}
\section{Asymptotics of $\overline{{\cal D}}_{n}(\mu_{0})$ under the benchmark model}

Our objective here is to obtain the limiting distribution of $\overline{{\cal D}}_{n}(\mu_{0})$ under the null hypothesis of no 
predictability. Our assumptions are collected under Assumption 1 below 
and consist of a collection of high level assumptions that are general enough to accommodate most environments commonly encountered in economics and finance applications.

\noindent
{\bf Assumptions 1.} (i) The $u_{t}'s$ form a martingale difference sequence (m.d.s.) with respect to the natural filtration and the sequence $\eta_{t}=u_{t+1}^{2}-E[u_{t+1}^{2}]$ satisfies $\sum_{t=k_{0}}^{k_{0}-1+[(n-k_{0})r]}\eta_{t}/\sqrt{n-k_{0}} \stackrel{d}\rightarrow \phi \ W(r)$ for $r \in [0,1]$ with $W(r)$ denoting a standard scalar Brownian Motion and 
$\phi^{2}=\sum_{s=-\infty}^{\infty}\gamma_{\eta}(s)>0$ for $\gamma_{\eta}(s)=E[\eta_{t}\eta_{t+s}]$.  
(ii) There is a $\hat{\phi}^{2}_{n}$ such that $\hat{\phi}^{2}_{n}\stackrel{p}\rightarrow \phi^{2} \in (0,\infty)$. 
(iii) Under the null hypothesis the forecast errors satisfy  $\sup_{r}|\sum_{t=k_{0}}^{k_{0}-1+[(n-k_{0})r]}(\hat{e}_{\ell,t+1|t}^{2}-u_{t+1}^{2})/\sqrt{n-k_{0}}|=o_{p}(1)$
$\forall \ell \in \{0,1,2,\ldots,p\}$. 
(iv) $\mu_{0}$ satisfies $\mu_{0} \in (0,1) \setminus \{1/2\}$. (v) The size of the pool of predictors $p$ is fixed throughout. 

Assumption 1(i) rules out the presence of serial correlation in the $u_{t}'s$ and requires the sequence of demeaned squared errors driving (\ref{eq:eq2}) to satisfy a suitable FCLT. We also note from the expression of $\phi^{2}$ that the $u_{t}'s$ while being serially uncorrelated could be either conditionally homoskedastic or conditionally hetereoskedastic. 
Taking the $u_{t}'s$ to be an m.d.s. with an ARCH type variance combined with mild existence of moments requirements would satisfy our environment in 1(i). 
Assumption 1(ii) requires that a consistent estimator of the long run variance associated with the $\eta_{t}$'s to be available. Letting $\hat{\eta}_{t}$ denote a generic estimator of the $\eta_{t}'s$ a trivial choice under conditional homoskedasticity 
would be $\hat{\phi}^{2}=\sum_{t}\hat{\eta}_{t}^{2}/(n-k_{0})$ while under conditional heteroskedasticity 
one may use a Newey-West type formulation as in Deng and Perron (2008).
As we are operating under the global null, such an estimator is readily available using the residuals from the benchmark model in (\ref{eq:eq2}). Alternative formulations could also 
be based on the $\hat{e}_{0,t+1}$'s or the $\hat{e}_{j,t+1}$'s as also justified by Assumption 1(iii). The key point to make at this stage is that these options will have no bearing on the limiting {\it null} distribution of $\overline{{\cal D}}_{n}(m_{0})$. However, such alternative choices may have an important 
influence on power, an issue we postpone to further below. Assumption 1(iii) can be viewed as a correct specification assumption in the sense that under the null hypothesis squared forecast errors are understood to behave like their 
true counterparts. Such a property holds within a broad range of contexts as established in 
Berenguer-Rico and Nielsen (2019), including settings with purely stationary or highly persistent predictors. 
Assumption 1(iv) imposes a minor restriction on $m_{0}=[(n-k_{0})\mu_{0}]$ used in the construction of $\overline{\cal D}_{n}(\mu_{0})$ to ensure that it has a non-degenerate asymptotic variance. 
To gain further intuition on this important point it is useful to explicitly evaluate the limiting variance, say $\omega^{2}$, of the numerator of (\ref{eq:eq3}) under the null hypothesis. Replacing $\hat{e}_{0,t+1}^{2}$ and $\hat{e}_{j,t+1}^{2}$ with $\eta_{t+1}=u_{t+1}^{2}-E[u_{t+1}^{2}]$ in (\ref{eq:eq3}), rearranging and taking expectations (see Lemma A1 in the appendix) results in 
\begin{eqnarray}
	\omega^{2} & = & \dfrac{(1-2\mu_{0})^{2}}{4 \mu_{0}(1-\mu_{0})} \phi^{2} \label{eq:eq6}
\end{eqnarray}
\noindent so that the availability of a consistent estimator for $\phi^{2}$ also ensures that ${\omega}^{2}$ can be estimated consistently provided that $\mu_{0}$ satisfies Assumption 1(iv). Expression (\ref{eq:eq6}) also highlights the well known variance degeneracy problem one would face in this context if we had instead used the full sample mean associated with the forecast errors of the benchmark model by setting $\mu_{0}=1/2$. Finally Assumption 1(v) 
draws attention to the fact that all our results rely on 
the assumption that the pool of predictors is fixed so that the asymptotics are taken solely as $n\rightarrow \infty$. Although this may appear as restrictive for a setting with many predictor candidates it is important to point out that our proposed tests can and do accommodate potentially large magnitudes of $p$ including for instance $p$ near $n$ as illustrated in our Monte-Carlo experiments further below. 

\noindent 
{\bf Proposition 1}. {\it Under the benchmark model in (\ref{eq:eq2}), assumptions 1(i)-(v) and as $n\rightarrow \infty$ we have \vspace{-0.8cm}
	\begin{eqnarray}
		\overline{{\cal D}}_{n}(\mu_{0}) & \stackrel{d}\rightarrow & {\cal Z} \label{eq:eq7}
	\end{eqnarray}	
	with ${\cal Z}$ denoting a standard normally distributed random variable. 
}
\vspace{0.1cm}

As it is customary in this literature 
(\ref{eq:eq7}) is implemented using one-sided (right tail) tests so that a rejection of the null provides support for the availability of at least one active predictor that helps generate more accurate forecasts than 
the benchmark model.  

\vspace{-0.4cm}
\section{Asymptotic Local Power Properties of $\overline{{\cal D}}_{n}(\mu_{0})$}

We next explore the ability of $\overline{{\cal D}}_{n}(\mu_{0})$ to detect predictability induced by one or more of the available $p$ predictors. Two aspects we are interested in exploring are the influence of the persistence properties of predictors on power and the role played by the choice of $\mu_{0}$ in $\overline{{\cal D}}_{n}(\mu_{0})$. We analyse local power within the following parameterisation \vspace{-0.2cm}
\begin{eqnarray}
	y_{t+1} & = & {\bm \beta}_{n}'{\bm x}_{t}+u_{t+1} \label{eq:eq8}
\end{eqnarray}
with ${\bm \beta}_{n}=n^{-\gamma} {\bm \beta}^{*}$ for ${\bm \beta}^{*}=(\beta_{1}^{*},\ldots,\beta_{p}^{*})'$,
noting that within (\ref{eq:eq8}) all but one of the $\beta_{i}^{*}$'s may be zero.
We let \vspace{-0.2cm}
\begin{equation}
	{\cal I}^{*} = \{1\leq j\leq p \colon \beta_{j}^{*}\neq 0\} \label{eq:eq9}
\end{equation} 
\noindent denote the set of active predictors with cardinality $|{\cal I}^{*}|=q\geq 1$ (i.e., the size of the true model).

In what follows we establish the local power properties of $\overline{{\cal D}}_{n}(\mu_{0})$ across three scenarios. In a first instance we take all $p$ components of ${\bm x}_{t}$ to be stationary and ergodic processes (scenario {\bf A}). We then focus on the case where the ${\bm x}_{t}$'s are parameterised as persistent processes (scenario {\bf B}). Finally our last scenario sets ${\bm x}_{t}=({\bm x}_{1,t},{\bm x}_{2,t})'$ with ${\bm x}_{1,t}$ and ${\bm x}_{2,t}$ containing non-persistent and persistent predictors respectively (scenario {\bf C}). In this latter case (\ref{eq:eq8}) takes the following form \vspace{-0.2cm}
\begin{eqnarray}
	y_{t+1} & = & {\bm \beta}_{1n}'{\bm x}_{1t}+{\bm \beta}_{2n}'{\bm x}_{2t}+u_{t+1} \label{eq:eq10}
\end{eqnarray}
with ${\bm x}_{1t}=(x_{1,t},\ldots,x_{p_{1},t})$ and ${\bm x}_{2t}=(x_{p_{1}+1,t},\ldots,x_{p,t})$ so that 
the pool of $p$ predictors is sub-divided into two types. The slope parameter vectors are in turn specified as ${\bm \beta}_{1n}=n^{-\gamma_{1}} {\bm \beta}_{1}^{*}$ for ${\bm \beta}_{1}^{*}=(\beta_{1,1}^{*},\ldots,\beta_{1,p_{1}}^{*})'$
and ${\bm \beta}_{2n}=n^{-\gamma_{2}} {\bm \beta}_{2}^{*}$ for ${\bm \beta}_{2}^{*}=(\beta_{2,p_{1}+1}^{*},\ldots,\beta_{2,p}^{*})'$. 
This mixed environment requires us to also modify the formulation of the active set of 
predictors included in the DGP. For this purpose we let \vspace{-0.2cm}
\begin{equation}
	{\cal I}_{1}^{*} = \{1\leq j\leq p_{1} \colon \beta_{1,j}^{*}\neq 0\} \label{eq:eq11} 
\end{equation} 
\begin{equation}
	{\cal I}_{2}^{*} = \{p_{1}+1\leq j\leq p \colon \beta_{2,j}^{*}\neq 0\} \label{eq:eq12}
\end{equation} 

\noindent with $|{\cal I}_{1}^{*}|=q_{1}$ and $|{\cal I}_{2}^{*}|=q_{2}$. In this setting the specification in (\ref{eq:eq10}) has $q_{1}$ active predictors satisfying scenario A and $q_{2}$ active predictors satisfying scenario B. 

Assumption 2A summarises our operating framework when all predictors are assumed to be purely stationary. 

\noindent
{\bf Assumption 2A.} (i) Assumptions 1(i), 1(ii) and 1(iv)-(v) hold. (ii) The model in (\ref{eq:eq8}) holds with $\gamma=1/4$. (iii) The $p$ predictors satisfy $\sup_{\lambda \in[0,1]}|\sum_{t=1}^{[n\lambda]} x_{it}x_{jt}/n - \lambda E[x_{it}x_{jt}]|=o_{p}(1)$ and $\sum_{t=1}^{[n\lambda]}x_{it}u_{t+1}/\sqrt{n}=O_{p}(1)$ for $i,j=1,\ldots,p$. 

Note that part (i) of Assumption 2A excludes 1(iii) as we are no longer operating under the 
null hypothesis. Part (ii) sets the rate at which we explore departures from the null. 
Here it is useful to point out that the local to the null parameterization with $\gamma=1/4$ as opposed to the more familiar $n^{1/2}$ rate is not in any way the result of our modelling environment or methods. This choice is essentially driven by the fact that we are conducting inferences using squared errors rather than their level. 
The remainder parts of Assumption 2A require that a uniform law of large number applies to the 
predictors and that a suitable CLT holds ensuring the uniform boundedness of relevant sample moments. Another important point to make here is that under the local to the null specification in (\ref{eq:eq8}) the residual variance estimated from the benchmark model will continue to remain consistent so that Assumption 1(ii) requiring $\hat{\phi}^{2} \stackrel{p}\rightarrow \phi^{2}$ continues to hold when estimated using the residuals from the benchmark model or any of the alternative choices mentioned earlier.  

Regarding the scenario with persistent predictors we parameterise these as mildly integrated processes via  \vspace{-0.28cm}
\begin{eqnarray}
	x_{jt} & = & \left(1-\dfrac{c_{j}}{n^{\alpha}}\right) x_{jt-1}+v_{jt} \ \ \ \ j=1,\ldots,p \label{eq:eq13}
\end{eqnarray}
\noindent
where $c_{j}>0$, $\alpha \in (0,1)$ and $v_{jt}$ denotes a random disturbance term.
The high level assumptions we impose under Assumption 2B explicitly accommodate dynamics such as (\ref{eq:eq13}) and follow directly from 
Phillips and Magdalinos (2009). We also let ${\bm \Sigma}_{vv}$ denote the $p\times p$ covariance of the 
$v_{jt}$'s and refer to its diagonal components as $\sigma^{2}_{v_{j}}$ and its off-diagonal 
components as $\sigma_{v_{i}v_{j}}$ respectively. 

\noindent
{\bf Assumption 2B.} (i) Assumptions 1(i), 1(ii) and 1(iv)-(v) hold. (ii) The model in (\ref{eq:eq8}) holds with $\gamma=(1+2\alpha)/4$ for $\alpha \in (0,1)$. (iii) The $p$ predictors follow the process in (\ref{eq:eq13}) and satisfy  $\sum_{t=1}^{[n\lambda]}x_{it}x_{jt}/n^{1+\alpha}\stackrel{p}\rightarrow \lambda \sigma_{v_{i}v_{j}}/(c_{i}+c_{j})$, 
$\sum_{t=1}^{[n\lambda]}x_{jt}^{2}/n^{1+\alpha}\stackrel{p}\rightarrow \lambda \sigma_{v_{j}}^{2}/(2c_{j})$ and 
$\sum_{t=1}^{[n\lambda]}x_{jt}u_{t+1}/n^{\frac{1+\alpha}{2}}=O_{p}(1)$ for $i,j=1,\ldots,p$.

In the context of our specification in (\ref{eq:eq8}), Assumption 2B(iii) is guaranteed to hold when the predictors follow the mildly integrated process in (\ref{eq:eq13}) as established in Lemmas 3.1-3.3 of Phillips and Magdalinos (2009). Our last assumption accommodates an environment that combines stationary and persistent predictors.

\noindent
{\bf Assumption 2C.} (i) Assumptions 1(i), 1(ii) and 1(iv)-(v) hold. (ii) The model in (\ref{eq:eq10}) holds with $\gamma_{1}=1/4$ and $\gamma_{2}=(1+2\alpha)/4$ for $\alpha \in (0,1)$. (iii) The pool of $p$ predictors consists of $p_{1}$ predictors satisfying Assumptions 2A(ii)-(iii) and $p_{2}=p-p_{1}$ predictors satisfying Assumptions 2B(ii)-(iii). 

\subsection{Local Power under Stationarity (scenario A)}

\noindent
{\bf Proposition 2A:} {\it Under Assumption 2A, $q \colon =|{\cal I}^{*}|$ active predictors in (\ref{eq:eq8}) with associated 
	slope parameters $\beta_{i}=n^{-1/4} \beta_{i}^{*}$ for $i \in {\cal I}^{*}$, and as $n\rightarrow \infty$ we have 
	\begin{eqnarray}
		\overline{{\cal D}}_{n}(\mu_{0}) & \stackrel{d}\rightarrow & {\cal Z}+ \ g(\mu_{0},\pi_{0},\phi)\ \ \frac{1}{p} \sum_{j=1}^{p}\left(
		\sum_{i \in {\cal I}^{*}}\beta_{i}^{*} \dfrac{E[x_{it}x_{jt}]}{\sqrt{E[x_{jt}^{2}]}} \right)^{2} \label{eq:eq14}
	\end{eqnarray}
	\noindent where \vspace{-0.32cm}
	\begin{eqnarray}
		g(\mu_{0},\pi_{0},\phi) & = & \dfrac{2 \sqrt{1-\pi_{0}}\sqrt{\mu_{0}(1-\mu_{0})}}{\sqrt{\phi^{2}}(1-2\mu_{0})} \label{eq:eq15}
	\end{eqnarray}
with $\pi_{0}$ denoting the sample fraction used to initiate the recursive forecasts and $\mu_{0}$ and $1-\mu_{0}$ the proportions used in the split sample averages in (\ref{eq:eq3}). 
	
}

The result in (\ref{eq:eq14}) establishes the consistency of our proposed test and its ability to detect departures from the constant mean model in (\ref{eq:eq2}) when predictors are taken to be stationary processes. The expression in (\ref{eq:eq14}) and its counterparts under persistence presented further below offer novel insights on the asymptotic behaviour of predictive accuracy comparisons not explored in the existing literature. We can also observe that power is monotonic as the non-centrality component in (\ref{eq:eq14}) is non-decreasing as the slope parameters of the active predictors increase. Another implication of
(\ref{eq:eq14}) is that under fixed alternatives $\overline{\cal D}_{n}(\mu_{0})\rightarrow \infty$ and more specifically \vspace{-0.28cm}
\begin{eqnarray}
	\overline{\cal D}_{n}(\mu_{0}) & \stackrel{H_{1}}{=} & O_{p}(\sqrt{n}). \label{eq:eq16}
\end{eqnarray}

\noindent 
{REMARK 2}. The local power result in Proposition 2A has been obtained under local departures from the null that are of order $n^{-1/4}$ 
rather than the conventional square root rates one typically observes in stationary environments. This is not due to the way the test statistic $\overline{D}_{n}(\mu_{0})$ has been constructed or to our inference framework in general. The main reason for operating under such a rate comes from the use of squared errors which result in the squaring of the relevant parameters in the DGP. 

To gain further intuition on the 
formulation of the second component in the right hand side of (\ref{eq:eq14}) it is useful to specialise the result to a single active predictor scenario. Suppose that there is a single active predictor, say $x_{at}$, with associated slope parameter $\beta_{an}=n^{-1/4}\beta_{a}^{*}$. It now follows directly from (\ref{eq:eq14}) that \vspace{-0.28cm}
\begin{eqnarray}
	\overline{{\cal D}}_{n}(\mu_{0}) & \stackrel{d}\rightarrow & {\cal Z}+g(\mu_{0},\pi_{0},\phi) \  (\beta_{a}^{*})^{2} \  E[x_{at}^{2}] \  \frac{1}{p}\sum_{j=1}^{p}\rho^{2}_{a,j}. \label{eq:eq17}
\end{eqnarray}

It is here interesting to note the role played by the correlation between 
the single predictor $x_{at}$ driving the DGP in (\ref{eq:eq8}) and the remaining components of the predictor pool (i.e. the irrelevant candidates). The higher this correlation is the stronger we expect power to be. This clearly conforms with intuition since the models are estimated with one predictor at a time. A particular fitted specification containing a predictor other than $x_{at}$ and therefore misspecified will nevertheless continue to dominate 
the intercept only model in an MSE sense provided that this pseudo-signal contains relevant information about $x_{at}$. Note also that this does not mean that in an environment where all predictors in the pool are uncorrelated with $x_{at}$ power will vanish as we have $\rho^{2}_{a,a}=1$ by construction, implying that the second component in the right hand side of (\ref{eq:eq17}) will always be strictly positive under our assumptions. Note however that in such instances where all candidate predictors are uncorrelated with $x_{at}$ the size of the predictor pool $p$ will have a detrimental impact on power, all other things kept equal. 

Another important implication of (\ref{eq:eq17}) is the favourable impact that the variance of $x_{at}$ has on power. The more persistent $x_{at}$ is, the better the power is expected to be. This hints at the fact that all other things being equal the presence of persistent predictors in the pool will improve the detection ability of our test. We can also note that the role of persistence may manifest itself not only via $E[x_{at}^{2}]$ but also via $\rho_{a,j}^{2}$ due to the spurious correlation phenomenon characterising persistent processes. These issues are explored in the next proposition.

\subsection{Local Power under Persistence (scenario B)}

\noindent
{\bf Proposition 2B:} {\it Under Assumption 2B, $q=|{\cal I}^{*}|$  active predictors in (\ref{eq:eq8}) with 
	slope parameters $\beta_{i}=n^{-(1+2\alpha)/4} \beta_{i}^{*}$ for $i \in {\cal I}^{*}$ and as $n\rightarrow \infty$ we have 
	\begin{eqnarray}
		\overline{{\cal D}}_{n}(\mu_{0}) & \stackrel{d}\rightarrow & {\cal Z}+ g(\mu_{0},\pi_{0},\phi) \  \dfrac{1}{p}\sum_{j=1}^{p} \left(
		\sum_{i \in {\cal I}^{*}}\beta_{i}^{*} \dfrac{\sigma_{v_{i} v_{j}}}{\sqrt{\sigma_{v_{j}}^{2}}} \sqrt{\dfrac{2c_{j}}{(c_{i}+c_{j})^{2}}} \right)^{2} \label{eq:eq18}
	\end{eqnarray}
	with $g(\mu_{0},\pi_{0},\phi)$ as in (\ref{eq:eq15}).}

The result in (\ref{eq:eq18}) highlights the beneficial impact that predictor persistence will have on the detection 
ability of $\overline{\cal D}_{n}(\mu_{0})$. This can also be observed by focusing on fixed alternatives 
under which we can immediately infer from (\ref{eq:eq18}) that \vspace{-0.28cm}
\begin{eqnarray}
	\overline{\cal D}_{n}(\mu_{0}) & \stackrel{H_{1}}{=} & O_{p}(n^{\frac{1+2\alpha}{2}}). \label{eq:eq19}
\end{eqnarray}
\noindent 
If we were to restrict all 
predictors to have the same non-centrality parameter, say $c_{i}=c$ $\forall i=1,\ldots,p$, (\ref{eq:eq18}) reduces to \vspace{-0.28cm}
\begin{eqnarray}
	\overline{{\cal D}}_{n}(\mu_{0}) & \stackrel{d}\rightarrow & {\cal Z}+ g(\mu_{0},\pi_{0},\phi) \  \dfrac{1}{p} \dfrac{1}{\sqrt{2c}} \sum_{j=1}^{p} \left(
	\sum_{i \in {\cal I}^{*}}\beta_{i}^{*} \dfrac{\sigma_{v_{i} v_{j}}}{\sqrt{\sigma_{v_{j}}^{2}}} \right)^{2} \label{eq:eq20}
\end{eqnarray}
\noindent which also suggests that all other things being equal power is expected to improve for smaller 
magnitudes of this non-centrality parameter. 

\subsection{Local Power under Mixed Predictors (scenario C)}

The pool of predictors now consists of $p_{1}$ purely stationary and $p_{2}$ persistent predictors with $p_{1}+p_{2}=p$ and we 
let ${\cal J}_{1}$ and ${\cal J}_{2}$ denote the sets associated with the stationary and persistent predictors respectively
so that $|{\cal J}_{1}|=p_{1}$ and $|{\cal J}_{2}|=p-p_{1}$. 

\noindent
{\bf Proposition 2C:} {\it Under Assumption 2C, $q_{1}=|{\cal I}^{*}_{1}|$ and $q_{2}=|{\cal I}^{*}_{2}|$ active predictors in (\ref{eq:eq10}) with 
	slope parameters $\beta_{1,i}=n^{-1/4} \beta_{1,i}^{*}$ for $i \in {\cal I}_{1}^{*}$ and $\beta_{2,i}=n^{-(1+2\alpha)/4} \beta_{2,i}^{*}$ for $i \in {\cal I}_{2}^{*}$ we have as $n\rightarrow \infty$ 	
{\footnotesize 
		\begin{align}
\overline{{\cal D}}_{n}(\mu_{0}) & \stackrel{d}\rightarrow  {\cal Z}+ \dfrac{g(\mu_{0},\pi_{0},\phi)}{p} 
			\left(
			\sum_{j \in {\cal J}_{1}}\left(\sum_{i \in {\cal I}_{1}^{*}} \beta_{i}^{*} \dfrac{E[x_{it}x_{jt}]}{\sqrt{E[x_{jt}^{2}]}}\right)^{2} +
			\sum_{j \in {\cal J}_{2}}\left(\sum_{i \in {\cal I}_{2}^{*}} \beta_{i}^{*} \dfrac{\sigma_{v_{i}v_{j}}}{\sqrt{\sigma^{2}_{v_{j}}}} \sqrt{\dfrac{2c_{j}}{(c_{i}+c_{j})^{2}}} \right)^{2}
			\right)
			\label{eq:eq21}
		\end{align}
	}
	with $g(\mu_{0},\pi_{0},\phi)$ as in (\ref{eq:eq15}).
}

\vspace{0.1cm}
\noindent
{REMARK 3.} The expressions in (\ref{eq:eq14}), (\ref{eq:eq18}), (\ref{eq:eq21}) provide useful insights on suitable choices of $\mu_{0}$ when constructing our test statistic. As $\mu_{0}$ affects local power via $g(\mu_{0},\pi_{0},\phi)$ a choice in the vicinity of 0.5 
is expected to lead to the most favourable power outcomes.
\vspace{-0.68cm}

\section{A Power enhancing modification of $\overline{\cal D}_{n}(\mu_{0})$}

\vspace{-0.2cm}
We here consider a modification of $\overline{D}_{n}(\mu_{0})$ designed to enhance its power without affecting its null distribution. Our proposal is in the spirit of Fan, Liao and Yao (2015) and involves augmenting the test statistic with a quantity that converges to 0 under the null while diverging under the alternative of at least one active 
predictor. For this purpose we introduce the quantity $d_{nj}=\sum_{t=k_{0}}^{n-1}(\hat{e}_{0,t+1|t}-\hat{e}_{j,t+1|t})^{2}/(n-k_{0})$. 
Within our nested context and under the null of the benchmark model we have $d_{nj}=O_{p}(n^{-1/2})$ $\forall j=1,\ldots,p$ while under the alternative whereby the true model contains at least one active predictor  we have $\widetilde{d}_{nj}\equiv \sqrt{n-k_{0}} \ d_{nj}/\hat{\omega}_{n}=O_{p}(1)$ $\forall j=1,\ldots,p$ with the associated limiting random variable being strictly positive. This prompts us to propose the following augmentation to 
$\overline{\cal D}_{n}(m_{0})$ \vspace{-0.2cm}
\begin{eqnarray}
	\overline{{\cal D}}_{n}^{d}(m_{0}) & = & \dfrac{1}{p} \sum_{j=1}^{p} ({\cal D}_{n}(m_{0},j)+\widetilde{d}_{nj}) \label{eq:eq22}
\end{eqnarray}
\noindent which we expect to be power enhancing while also being size-neutral. Noting that \vspace{-0.32cm}
\begin{eqnarray}
	\overline{\cal D}_{n}^{d}(m_{0})-\overline{\cal D}_{n}(m_{0}) & = & \dfrac{1}{p}\sum_{j=1}^{p}\widetilde{d}_{nj} \label{eq:eq23}
\end{eqnarray}
\noindent Proposition 3 below formalises these observations. 

\noindent
{\bf Proposition 3.} {\it  (i) Under Assumptions 1(i)-(v) and the null hypothesis we have 
	$\overline{\cal D}_{n}^{d}(m_{0})-\overline{\cal D}_{n}(m_{0}) \stackrel{p}\rightarrow 0$ as $n\rightarrow \infty$. (ii) Under Assumptions 2A, 2B or 2C we have 
	$\overline{\cal D}_{n}^{d}(m_{0})-\overline{\cal D}_{n}(m_{0}) \stackrel{p}\rightarrow Q_{\ell}>0$, $\ell=A,B,C$ with $Q_{\ell}$ given by the second component in the right hand side of (\ref{eq:eq14}), (\ref{eq:eq18}) and (\ref{eq:eq21}) respectively.} 

A key implication of Proposition 3 (ii) is that a test of size 
$\alpha$ based on ${\cal D}_{n}^{d}(\mu_{0})$
will be strictly preferable in terms of local power to a test of the same size based on ${\cal D}_{n}(\mu_{0})$. A more formal comparison using Pitman's asymptotic relative efficiency is also informative as it takes a particularly simple form in the present context. Indeed, our local power 
results in (\ref{eq:eq14}), (\ref{eq:eq18}), (\ref{eq:eq21}) combined with Proposition 3(ii) above  imply that \vspace{-0.2cm}
\begin{eqnarray}
	\text{ARE}({\cal D}_{n}(\mu_{0}),{\cal D}_{n}^{d}(\mu_{0})) & = & 1/2<1 \label{eq:eq24}
\end{eqnarray}
\noindent
irrespective of any model specific parameters.

\noindent
REMARK 4: The above power enhancing tranformation based on $\widetilde{d}_{nj}$ is analogous to implementing an adjustment to the forecast errors associated with the larger models. 
More specifically, implementing our main test statistic in (\ref{eq:eq3}) with the $\hat{e}_{j,t+1}^{2}$'s replaced with say $\widetilde{e}_{j,t+1}^{2}=\hat{e}_{j,t+1}^{2}-(\hat{e}_{0,t+1}-\hat{e}_{j,t+1})^{2}$ results in a formulation that is algebraically identical to (\ref{eq:eq22}). It is now interesting to observe that these $\widetilde{e}_{j,t+1}^{2}$'s are essentially adjusting the 
$\hat{e}_{j,t+1}^{2}$'s for estimation noise coming from the estimation of the larger model when its true parameters are zero. This is precisely the motivation behind the well known Clark and West adjustment to equal predictive accuracy tests proposed in Clark and West (2007).  Unlike the setting in Clark and West (2007) however our proposed test statistics result in formal normal limits rather than approximate ones.

\vspace{-0.6cm}
\section{Detecting the key player}

Upon rejection of the benchmark model it becomes interesting to explore ways of identifying the predictors driving these departures from the null hypothesis. In this context we distinguish between two settings and 
obtain the corresponding limiting behaviour of $
\hat{j}_{n} \in \arg \max_{j=1,\ldots,p} {\cal D}_{n}(m_{0},j)$ and $
\hat{j}_{n}^{d} \in \arg \max_{j=1,\ldots,p} {\cal D}_{n}^{d}(m_{0},j)$ which select the predictor that 
results in the greatest MSE spread relative to the benchmark model. 

In a first instance we evaluate the large sample behaviour of these estimators when the DGP contains a single active predictor (i.e. $q=|{\cal I}^{*}|=1$ in (\ref{eq:eq9})) that can be either stationary or persistent. We subsequently extend our analysis to environments with multiple predictors that are again assumed to be of the same type in their persistence properties (i.e. all stationary or all persistent). Finally we consider the case of mixed predictors as in (\ref{eq:eq11})-(\ref{eq:eq12}) with the joint presence of stationary and persistent active predictors numbering $q_{1}$ and $q_{2}$ respectively. 
The large sample behaviour of these key player estimators is summarised in the following Proposition. 

\noindent
{\bf Proposition 4}. {\it (i) Under Assumptions 2A or 2B and as $n\rightarrow \infty$ we have $\{\hat{j}_{n},\hat{j}_{n}^{d}\}\stackrel{p}  \rightarrow j_{0} \in {\cal I}^{*}$ for $q\geq 1$. (ii) Under Assumptions 2C and as $n\rightarrow \infty$ we have $\{\hat{j}_{n},\hat{j}_{n}^{d}\}\stackrel{p}\rightarrow j_{0} \in {\cal I}_{1}^{*}\cup {\cal I}_{2}^{*}$.} 
\vspace{0.2cm}

When the DGP consists solely of a single predictor (stationary or persistent), part (i) of Proposition 4 implies that $\hat{j}_{n}$ or $\hat{j}_{n}^{d}$ will be consistent for that true predictor asymptotically. When there are multiple predictors of the same type the same result implies that $\hat{j}_{n}$ or $\hat{j}_{n}^{d}$ remain consistent for one of the $q>1$ active predictors i.e. $\hat{j}_{n}$ or $\hat{j}_{n}^{d}$ is consistent for one of the true components in ${\cal I}^{*}$. Part (ii) of Proposition 4 relates to a scenario with mixed active predictors and states that in such a mixed setting $\hat{j}_{n}$ or $\hat{j}_{n}^{d}$ will continue to point to one of the true
predictors which may come from either of the two sets. 

Using the results provided in the proof of Proposition 4 it is useful to illustrate the mixed predictor scenario via a simple 
example of a predictive regression with two active predictors, say $y_{t+1}=\theta_{0}+\beta_{an}x_{at}+\beta_{bn}x_{bt}+u_{t+1}$ with 
$x_{at} \in {\cal I}_{1}^{*}$, $x_{bt}\in {\cal I}_{2}^{*}$ and as before $\beta_{an}=\beta_{a}^{*}/n^{1/4}$ and 
$\beta_{bn}=\beta_{b}^{*}/n^{(1+2\alpha)/4}$. Proposition 4(ii) clearly applies and implies that $\hat{j}_{n}$ or $\hat{j}_{n}^{d}$ will asymptotically point to either $x_{at}$ or $x_{bt}$. More specifically (see proof of Proposition 4),   
we have that $\hat{j}_{n}$ or $\hat{j}_{n}^{d}$ will asymptotically point to $x_{bt}$ (the persistent predictor)
if \vspace{-0.28cm}
\begin{eqnarray}
	(\beta_{b}^{*})^{2} & > & \dfrac{E[x_{at}^{2}]}{(\sigma^{2}_{v_{b}}/2c_{b})} (\beta_{a}^{*})^{2} \label{eq:eq25}
\end{eqnarray}
and to $x_{at}$ otherwise. It is now interesting to observe from (\ref{eq:eq25}) that $\hat{j}_{n}$ or $\hat{j}_{n}^{d}$ is expected to pick $x_{bt}$ when the squared slope associated with this predictor exceeds the {\it scaled} slope of $x_{at}$ with the scaling factor given by the ratio of the 
variances of the two predictors. As the ratio of these variances is likely to be small due the higher persistence of  
$x_{bt}$ the procedure is also more likely to identify the persistent predictor unless the slope associated with  
$x_{at}$ is particularly large relative to that of $x_{bt}$. More generally, these results suggest that the key player selected by our proposed methods will depend on the relative magnitude of its associated slope combined with its relative variance (relative to the remaining active predictors). Thus when it comes to where $\hat{j}_{n}$ or $\hat{j}_{n}^{d}$ point asymptotically, there will be a trade-off between slope strength and variance dominance. 

\vspace{-0.68cm}
\section{Implementation and Experimental Properties}

This section aims to document the empirical properties of our proposed test and key player estimator in finite samples. Given our theoretical analysis about the superior local power properties of ${\cal D}_{n}^{d}(\mu_{0})$ versus ${\cal D}_{n}(\mu_{0})$ we concentrate our discussion on  ${\cal D}_{n}^{d}(\mu_{0})$  which we then follow with experiments documenting the correct decision frequencies associated with the proposed key player estimators. A {\sf supplementary appendix} accompanying this paper provides additional 
simulation based illustrations of the finite sample behaviour of our proposed tests.  

\noindent
\textit{\textbf{Implementation}}: The implementation of our test statistics in (\ref{eq:eq3}) and (\ref{eq:eq4}) requires the availability of an estimator of the long run variance associated with the numerator of (\ref{eq:eq3}) which we generically referred to as $\hat{\omega}_{n}^{2}$. The asymptotic outcomes documented above operated under the assumption that an estimator satisfying $\hat{\omega}^{2}_{n}\stackrel{p}\rightarrow \omega^{2}$ was available. Given the expression of $\omega^{2}$ obtained in (\ref{eq:eq6}) and the fact that we operate under given $\mu_{0}$ it is also clear that a consistent estimator of $\phi^{2}$ would also ensure the availability of a consistent estimator of $\hat{\omega}^{2}_{n}$. 

We may consider two alternative estimators of the long run variance in (\ref{eq:eq3}) using either residuals from the null model or the marginal regressions considered under the alternative. From  Assumption 1(iii) it is straightforward to note that these will be asymptotically equivalent under the null hypothesis of interest but they may result in potentially important differences in finite samples, when it comes to power in particular. For conditionally homoskedastic $\eta_{t}'s$ we consider \vspace{-0.2cm}
\begin{eqnarray}
	\hat{\omega}^{2,a}_{n} & = & \dfrac{(1-2\mu_{0})^{2}}{4\mu_{0}(1-\mu_{0})} 
	\dfrac{\sum_{t=k0}^{n-1} (\hat{e}_{0,t+1}^{2}-\overline{\hat{e}_{0}^{2}})^{2}}{n-k_{0}} \label{eq:eq26}, \\
	\hat{\omega}^{2,b}_{n,j} & = & \dfrac{(1-2\mu_{0})^{2}}{4\mu_{0}(1-\mu_{0})} 
	\dfrac{\sum_{t=k0}^{n-1} (\hat{e}_{j,t+1}^{2}-\overline{\hat{e}_{j}^{2}})^{2}}{n-k_{0}}. \label{eq:eq27}
\end{eqnarray}

\noindent
which can also be readily adapted to accommodate conditionally heteroskedastic $u_{t}'s$ (equivalently, serial correlation in the $\eta_{t}'s$) using Newey-West type formulations. Letting $\hat{\eta}_{t}$ denote a generic estimator of 
$\eta_{t}$, such an estimator for the counterpart to $\hat{\omega}^{2,a}_{n}$ above would be given by
\vspace{-0.2cm} 
\begin{eqnarray}
	\widetilde{\omega}_{n}^{2,a} & = & \dfrac{(1-2\mu_{0})^{2}}{4\mu_{0}(1-\mu_{0})} \
	\sum_{s=-m}^{m}\left(1-\left|\dfrac{s}{n-k_{0}}\right|\right)\hat{\gamma}_{\eta}(s) \label{eq:eq28}
\end{eqnarray}
\noindent where $\hat{\gamma}_{\eta}(s)=\sum \hat{\eta}_{t}\hat{\eta}_{t-s}/n-k_{0}$ for 
$\hat{\eta}_{t}=\hat{u}_{0,t+1}^{2}-\hat{\sigma}^{2}_{0,u}$ and similarly for (\ref{eq:eq27}). Note that (\ref{eq:eq28}) specialises to (\ref{eq:eq26}) under conditional homoskedasticity, whereby $\gamma_{\eta}(s)=0$ $\forall s \neq 0$ and $m$ refers to a suitable bandwidth for which one may consider the rule of thumb $m=m_{n}=0.75 (n-k_{0})^{1/3}$. In the simulations presented below we analyse finite sample size and power using the variance normaliser in (\ref{eq:eq27}). 

\noindent
\textit{\textbf{Empirical Size}}: The DGP is given by the benchmark specification in (\ref{eq:eq2}) with $\theta_{0}$ set as equal to one throughout. The pool of $p$ predictors is taken to follow the VAR(1) process ${\bm x}_{t}={\bm \Phi} {\bm x}_{t-1}+{\bm v}_{t}$ with ${\bm v}_{t} \sim N(0,{\bm \Sigma_{vv}})$ which we parameterise in ways that can distinguish between uncorrelated, weakly correlated and strongly correlated predictors. 
We consider the following scenarios for the persistence properties of the predictors: {(A)} ${\bm \Phi}=0.50 \ {\bm I}_{p}$, {(B)} ${\bm \Phi}=0.95 \ {\bm I}_{p}$ and {(C)} ${\bm \Phi}=({\bm \Phi}_{1},{\bm \Phi}_{2})$, ${\bm \Phi}_{1}=0.5\ {\bm I}_{p_{1}}$, ${\bm \Phi}_{2}=0.95 \ {\bm I}_{p-p_{1}}$. Letting ${\bm \Omega}$ denote the covariance matrix of $(u_{t},{\bm v}_{t})'$ we write 

\vspace{-0.6cm}
\begin{eqnarray}
	{\bm \Omega}_{p+1 \times p+1} & = & 
	\begin{pmatrix}
		\sigma^{2}_{u} & {\bm \sigma}_{uv}' \\
		{\bm \sigma}_{uv} & {\bm \Sigma}_{vv}
	\end{pmatrix}
	\label{eq:eq29}
\end{eqnarray}
where ${\bm \sigma}_{uv}=(\sigma_{uv_{1}},\sigma_{uv_{2}},\ldots,\sigma_{uv_{p}})'$ collects the 
covariances between the shocks to $y_{t}$ and the shocks to individual predictors and ${\bm \Sigma}_{vv}$ is the $p\times p$ covariance matrix of the $p$ predictors. 
Our experiments involving either purely stationary, purely persistent or mixed predictors are conducted across three configurations of ${\bm \Omega}$: (i) ${\bm \Omega}_{0}$: $\sigma^{2}_{u}=1, {\bm \sigma}_{uv}={\bm 0}_{p\times 1}, {\bm \Sigma}_{vv}={\bm I}_{p}$, (ii) ${\bm \Omega}_{1}$: $\sigma^{2}_{u}=1, {\bm \sigma}_{uv}={\bm 0}_{p\times 1}, {\bm \Sigma}_{vv}=[0.5^{|i-j|}]_{i,j}$ and (iii) ${\bm \Omega}_{2}$: $\sigma^{2}_{u}=1, {\bm \sigma}_{uv}=[(-0.5)^{j}]_{j}, {\bm \Sigma}_{vv}=[0.5^{|i-j|}]_{i,j}$.
Accordingly we label these size related DGPs as {\bf (A-i)-(A-iii), (B-i)-(B-iii)} and {\bf (C-i)-(C-iii)}.
Scenario (i) forces all $p$ predictors to be uncorrelated between themselves. It also requires the shocks to the predictors and the shocks to the predictand to be uncorrelated as does scenario (ii). In this latter case predictors are now allowed to be correlated. Finally scenario (iii) allows for the shocks to predictand and predictors to be contemporaneously correlated. 

Empirical size outcomes are obtained for $p \in \{10,50,500\}$ and samples of 
size $n=500$ with $\pi_{0}=0.25$ used as the starting point for generating recursive forecasts i.e. $n-k_{0}=375$. For the sample-split location of our test statistic we experiment with $\mu_{0} \in \{0.35,0.40,0.45\}$. Results are collected in Table 1 below using 5000 Monte-Carlo replications and a nominal size of 10\%. It is worth pointing out that the 
case $p=500$ implies an environment where the number of predictors exceeds the effective sample size of $n-k_{0}=375$. Recalling that our main result in Proposition 1 is obtained under $n\rightarrow \infty$, our chosen parameterisations of the pair $(n,p)$ is meant to illustrate the finite sample adequacy of our asymptotics even when $n$ lies below $p$ or is near $p$. 

From Table 1 we can note that the {\it one predictor at a time} 
approach based on ${\cal D}_{n}^{d}(\mu_{0})$ appears to be robust to the dimension of the predictor pool with almost identical size estimates obtained across $p=10$, $p=50$ and $p=500$ predictors. 
This highlights the excellent approximation provided by our asymptotics even when $n-k_{0}$ is smaller than $p$. Outcomes can also be seen to be robust to predictor
persistence as expected from our result in Proposition 1. 
Equally importantly, we can highlight the fact that the chosen sample split location $\mu_{0}$ has very little influence on outcomes with almost identical empirical sizes obtained across all chosen magnitudes of $\mu_{0}$. This is particularly important as our earlier local power analysis suggested that choosing $\mu_{0}$ in the vicinity of 0.5 should result in better power outcomes, all other things being equal.  

\begin{table}[h]
	\centering
	\caption{Empirical Size of ${\cal D}_{n}^{d}(\mu_{0})$ (10\% Nominal)} \vspace{0.2cm}
\scalebox{0.8}{\begin{tabular}{lccccccccc}
		$\mu_{0}$ & p=10  & p=50  & p=500 & p=10  & p=50  & p=500 & p=10  & p=50  & p=500 \\ \hline
		& \multicolumn{3}{c}{A(i)} & \multicolumn{3}{c}{A(ii)} & \multicolumn{3}{c}{A(iii)} \\
		0.35  & 0.106 & 0.102 & 0.103 & 0.107 & 0.103 & 0.103 & 0.105 & 0.103 & 0.103 \\
		0.40  & 0.108 & 0.103 & 0.104 & 0.109 & 0.105 & 0.106 & 0.108 & 0.105 & 0.106 \\
		0.45  & 0.109 & 0.093 & 0.100 & 0.115 & 0.093 & 0.100 & 0.114 & 0.094 & 0.100 \\
		& \multicolumn{3}{c}{B(i)} & \multicolumn{3}{c}{B(ii)} & \multicolumn{3}{c}{B(iii)} \\
		0.35  & 0.105 & 0.103 & 0.103 & 0.106 & 0.102 & 0.103 & 0.107 & 0.102 & 0.103 \\
		0.40  & 0.114 & 0.105 & 0.105 & 0.114 & 0.104 & 0.104 & 0.114 & 0.104 & 0.104 \\
		0.45  & 0.116 & 0.099 & 0.101 & 0.119 & 0.100 & 0.103 & 0.122 & 0.100 & 0.103 \\
		& \multicolumn{3}{c}{C(i)} & \multicolumn{3}{c}{C(ii)} & \multicolumn{3}{c}{C(iii)} \\
		0.35  & 0.106 & 0.103 & 0.103 & 0.105 & 0.102 & 0.102 & 0.105 & 0.102 & 0.102 \\
		0.40  & 0.110 & 0.104 & 0.103 & 0.109 & 0.106 & 0.103 & 0.109 & 0.106 & 0.103 \\
		0.45  & 0.115 & 0.093 & 0.100 & 0.119 & 0.098 & 0.102 & 0.118 & 0.099 & 0.102 \\
	\end{tabular}}
	\label{tab:Tab1}
\end{table}


\noindent
\textit{\textbf{Empirical Power}}: We consider predictive regressions with up to four active predictors parameterised as \vspace{-0.4cm}
\begin{eqnarray}
	y_{t+1} & = & \theta_{0}+\beta_{an} x_{a,t}+\beta_{bn} x_{b,t}+\beta_{cn} x_{c,t}+\beta_{dn} x_{d,t}+u_{t+1} \label{eq:eq30}
\end{eqnarray}
with $\beta_{an}=\beta_{a}^{*}/n^{0.25}$, $\beta_{bn}=\beta_{b}^{*}/n^{0.25}$, $\beta_{cn}=\beta_{c}^{*}/n^{0.675}$,  and $\beta_{dn}=\beta_{d}^{*}/n^{0.675}$. The two predictors labelled as $\{a,b\}$ are chosen to be non-persistent while the predictors labelled as $\{c,d\}$ will have more persistence. Accordingly it will be understood that the pool of predictors to which $\{x_{at},x_{bt},x_{ct},x_{dt}\}$ belong is generated from a VAR(1) parameterised as in (C-iii) above. The active predictors $\{x_{at},x_{bt}\}$ belong to the first set of $p_{1}$ predictors and $\{x_{ct},x_{dt}\}$ belong to the second set of $p-p_{1}$ predictors. We take $x_{at}=x_{1t}$, $x_{bt}=x_{2t}$ and $x_{ct}=x_{p_{1}+1,t}$, $x_{dt}=x_{p_{1}+2,t}$. 
The {\it local} parameterisations of the slope parameters are chosen in a way to be compatible with our earlier local power analysis where we documented local departures of $n^{-0.25}$ under the stationary setting and $n^{-(1+2 \alpha)/4}$ under mild integratedness. For this latter case, setting $\alpha=0.85$ in the mildly integrated process in (\ref{eq:eq3}) results in $(1+2\alpha)/4=0.675$. As these persistent predictors are driven by the VAR(1) component with slopes 0.95 we may also note that this corresponds roughly to $(1-c/n^{\alpha})=(1-10/500^{0.85})\approx 0.95$. 

We consider 3 DGP configurations: {\bf (i)} $\beta_{a}^{*} \in \{2,3,4,5\}$, $\beta_{b}^{*} \in \{5,6,7,8\}$, $\beta_{c}^{*}={0}$, $\beta_{d}^{*}={0}$, {\bf (ii-a)}
$\beta_{a}^{*}={0}$, $\beta_{b}^{*}={0}$, $\beta_{c}^{*} \in \{2,3,4,5\}$, $\beta_{d}^{*} \in \{5,6,7,8\}$ and {\bf (ii-b)} $\beta_{a}^{*}={0}$, $\beta_{b}^{*}={0}$, $\beta_{c}^{*} \in \{5,6,7,8\}$, $\beta_{d}^{*} \in \{8,9,10,11\}$. 
Scenario (i) involves two active predictors $x_{at}$ and $x_{bt}$ that are not persistent (i.e., selected from the pool of $p_{1}$ predictors that follow autoregressive processes with slopes equal to 0.5). Scenarios (ii-a) and (ii-b) involve two active predictors $x_{ct}$ and $x_{dt}$ selected from the remaining pool of $p-p_{1}$ persistent predictors. Relative to (ii-a), DGP (ii-b) is characterised by a stronger signal to noise ratio. 

It is also 
useful to point out that with $n=500$ the chosen slope parameterisations translate into 
$\beta_{an}\in \{0.422,0.634, 0.845, 1.057\}$ and
$\beta_{bn}\in \{1.057, 1.269, 1.480, 1.692\}$ for scenario (i), $\beta_{cn}\in \{0.030, 0.045, 0.060, 0.075\}$ and 
$\beta_{dn}\in \{0.075, 0.090, 0.106,0.121\}$ for scenario (ii-a), $\beta_{cn}\in \{0.075, 0.090, 0.106, 0.121\}$ and 
$\beta_{dn}\in \{0.121, 0.136, 0.151, 0.166\}$ for scenario (ii-b). These highlight the fact that power is evaluated as the DGP moves further away from the null under a fixed sample size set at $n=500$. We also note that the above slope magnitudes span low, medium and high signal to noise ratios. 

Our power experiments are implemented using a predictor pool of $p=100$ predictors with the first half consisting of autoregressive processes with slopes set at 0.50 and the second half having slopes of 0.95. All experiments are
implemented using the covariance structure labelled as ${\bm \Omega}_{2}$ above. Outcomes associated with 
${\cal D}_{n}^{d}(\mu_{0})$ are collected in Table 2.

We note that power increases towards 100\% as the slope parameters move away from the null (each column of Table 2 corresponds to one pair of slopes, and moving rightwards along the table illustrates power progression for larger departures from the null). Under DGP(i) empirical powers lie in the vicinity of 100\% for $\mu_{0}\geq 0.40$ and across all slope parameterisations. DGP(ii-a) is associated with much weaker signal to noise ratios which as expected translate into much less favourable empirical powers. Nevertheless we do note powers as high as 75\% even in this context. Here it is important to recall that this DGP consists solely of persistent predictors. Had we used the slope magnitudes $(\beta_{an},\beta_{bn})$ instead of 
$(\beta_{cn},\beta_{dn})$ in this context all empirical powers would have resulted in 100\% or nearly 100\% correct decision frequencies. This can also be inferred from the outcomes based on DGP (ii-b) which uses larger slope magnitudes 
for the same persistent predictor scenario. Under $(\beta_{cn},\beta_{dn})=(0.121,0.166)$ which are much lower than 
the most favourable slope pairs considered in DGP(i) for instance we note empirical power outcomes in excess of 90\%.   

\begin{table}[h]
	\centering
	\caption{Empirical Power of ${\cal D}_{n}^{d}(\mu_{0})$ under DGPs (i)-(ii)} \vspace{0.5cm}
\scalebox{0.8}{\begin{tabular}{lcccc} \hline
		& \multicolumn{4}{c}{DGP (i)} \\
		$\beta_{an}$ & 0.423 & 0.634 & 0.846 & 1.057 \\
		$\beta_{bn}$ & 1.057 & 1.269 & 1.480 & 1.692 \\ \hline
		$\mu_{0}=0.35$ & 0.829 & 0.936 & 0.976 & 0.991 \\
		$\mu_{0}=0.40$ & 0.977 & 0.997 & 1.000 & 1.000 \\
		$\mu_{0}=0.45$ & 1.000 & 1.000 & 1.000 & 1.000 \\ \hline 
		& \multicolumn{4}{c}{DGP (ii-a)} \\
		$\beta_{cn}$ & 0.030 & 0.045 & 0.060 & 0.075 \\
		$\beta_{dn}$ & 0.075 & 0.090 & 0.106 & 0.121 \\ \hline
		$\mu_{0}=0.35$ & 0.174 & 0.231 & 0.304 & 0.346 \\
		$\mu_{0}=0.40$ & 0.217 & 0.296 & 0.391 & 0.464 \\
		$\mu_{0}=0.45$ & 0.349 & 0.500 & 0.641 & 0.747 \\ \hline 
		& \multicolumn{4}{c}{DGP (ii-b)} \\
		$\beta_{cn}$ & 0.075 & 0.090 & 0.106 & 0.121 \\
		$\beta_{dn}$ & 0.121 & 0.136 & 0.151 & 0.166 \\ \hline
		$\mu_{0}=0.35$ & 0.337 & 0.399 & 0.464 & 0.492 \\
		$\mu_{0}=0.40$ & 0.455 & 0.533 & 0.604 & 0.655 \\
		$\mu_{0}=0.45$ & 0.742 & 0.827 & 0.885 & 0.924 \\
	\end{tabular}}
	\label{tab:Tab2}
\end{table}

\vspace{0.1cm}

\noindent
\textit{\textbf{Finite Sample Properties of the key player estimator}}: In this last set of experiments we illustrate the result in Proposition 3 by documenting the behaviour of
the proposed {\it key player} estimator as $n$ is allowed to grow.
We base our evaluation of $\hat{j}_{n}^{d}$ on its ability to point to one of the true underlying active predictors. Although this estimator would be typically implemented upon rejection of the null hypothesis our empirical results below are understood to be unconditional in the sense that the correct decision frequencies associated with $\hat{j}_{n}^{d}$'s are 
averaged across all realisations rather than the ones associated with rejections of the null. This is justified here on the basis that $\hat{j}_{n}^{d}$ may still point to the true predictor (or one of the true predictors) even if the underlying inferences based on ${\cal D}_{n}^{d}(\mu_{0})$ do not result in a rejection of the null. Differently put, the realised magnitude of ${\cal D}_{n}^{d}(\mu_{0})$ could still be the largest amongst the $j=1,\ldots,p$ realisations even if it falls below the null hypothesis rejection threshold. 

The DGPs parallel the specifications labelled as (i)-(ii) in our earlier power analysis. The number of predictors is set at $p=100$ throughout. The specific parameterisations are: (i) $(\beta_{an},\beta_{bn})=(0.634, 1.269)$, $x_{at}=x_{1t}$, $x_{bt}=x_{2t}$, (ii-a) $(\beta_{cn},\beta_{dn})=(0.075,0.121)$, $x_{ct}=x_{51,t}$, $x_{dt}=x_{52,t}$, (ii-b) $(\beta_{cn},\beta_{dn})=(0.106,0.151)$, $x_{ct}=x_{51,t}$, $x_{dt}=x_{52,t}$. We implement the above experiments across samples of size $n=100$ and $n=200$. In DGP(i) the active predictors are $j_{0}\in \{1,2\}$. In DGPs (ii-a) and (ii-b) we have $j_{0}\in \{51,52\}$. Results are collected in Table 3 which displays relevant correct decision frequencies. An important primary observation we can make is the similarity of outcomes across the different sample split locations (i.e., choices for $\mu_{0}$) whose choice clearly does not matter much. For DGP (i) our estimator picks up $j_{0}=2$ in excess of 99\% of the times even under $(n=100,p=100)$. The preference for $j_{0}=2$ over $j_{0}=1$ which is also included as an active predictor is due to the much larger slope associated with $x_{2t}$ combined with the fact that these predictors have the same variance. For DGP (ii-a) which is driven by the two active predictors $x_{51,t}$ and $x_{52,t}$ we note that 
the bulk of $\hat{j}_{n}^{d}$'s decision frequencies converge towards $x_{52,t}$ (e.g., about 70\% under $n=200$). This can again be explained by the fact that $x_{52,t}$ has a stronger signal due to its larger slope parameter. Similar outcomes also characterise DGP (ii-b).

\begin{table}[htbp]
	\centering
\caption{Key Player Estimators: Frequency of detection of active predictors} \vspace{0.4cm}
\scalebox{0.8}{\begin{tabular}{lcccccc} \hline
		& \multicolumn{1}{l}{$\mu_{0}=0.35$} & \multicolumn{1}{l}{$\mu_{0}=0.40$} & \multicolumn{1}{l}{$\mu_{0}=0.45$} & \multicolumn{1}{l}{$\mu_{0}=0.35$} & \multicolumn{1}{l}{$\mu_{0}=0.40$} & \multicolumn{1}{l}{$\mu_{0}=0.45$} \\ \hline 
		& \multicolumn{3}{c}{n=100} & \multicolumn{3}{c}{n=200} \\ \hline 
		& \multicolumn{6}{c}{$DGP-(i)$, $j_{0}\in \{1,2\}$} \\ 
		$\hat{j}_{n}^{d}=1$ & 0.009 & 0.009 & 0.009 & 0.000 & 0.000 & 0.000 \\
		$\hat{j}_{n}^{d}=2$ & 0.991 & 0.991 & 0.991 & 1.000 & 1.000 & 1.000 \\
		& \multicolumn{6}{c}{$DGP-(ii-a)$, $j_{0}\in \{51,52\}$} \\
		$\hat{j}_{n}^{d}=51$ & 0.184 & 0.182 & 0.183 & 0.235 & 0.235 & 0.234 \\
		$\hat{j}_{n}^{d}=52$ & 0.413 & 0.414 & 0.413 & 0.703 & 0.702 & 0.703 \\
		$\hat{j}_{n}^{d} \neq \{51,52\}$ & 0.403 & 0.404 & 0.404 & 0.062 & 0.063 & 0.063 \\
		& \multicolumn{6}{c}{$DGP-(ii-b)$, $j_{0}\in \{51,52\}$} \\
		$\hat{j}_{n}^{d}=51$ & 0.252 & 0.253 & 0.252 & 0.270 & 0.271 & 0.270 \\
		$\hat{j}_{n}^{d}=52$ & 0.508 & 0.508 & 0.508 & 0.705 & 0.705 & 0.705 \\
		$\hat{j}_{n}^{d} \neq \{51,52\}$ & 0.239 & 0.239 & 0.239 & 0.025 & 0.025 & 0.025 
	\end{tabular}}
\label{tab:Tab4}
\end{table}

\vspace{-0.8cm}

\section{Application: Predictability of Economic Activity}

We apply our methods to the predictability of US economic activity and the monthly growth rate in industrial production in particular. The predictor pool consists of 130 lagged monthly series drawn from the FRED-MD database whose detailed constituents are discussed in McCracken and Ng (2016). These series have also been transformed and outliers processed as documented in McCracken and Ng (2016). 
FRED-MD consists of eight groups of time series and is closely aligned with the early Stock and Watson dataset (Stock and Watson (2002)): (1) output and income, (2) labor market, (3) housing, (4) consumption, (5) money and credit, (6) interest and exchange rates, (7) prices and (8) stock-market. Our selection of 130 predictors follows Giannone, Lenza and Primiceri (2021, GLP2021) using the data vintage provided by the authors and the same sample range of February 1960 to December 2014. This allows us to make comparisons between the inferences developed in this paper and existing findings in the literature. Recalling that our forecasts are generated recursively we set the starting point of the first recursion at the $165^{th}$ month (i.e., $k_{0}=[658 (0.25)]=165$). We implement our inferences across $\mu_{0} \in \{0.30, 0.35, 0.40, 0.45\}$ and consider both the ``raw" ${\cal D}_{n}(\mu_{0})$ statistic and its power enhanced counterpart ${\cal D}_{n}^{d}(\mu_{0})$. We also implement these two test statistic formulations using variance normalisers based on residuals obtained under the null and the alternative. Accordingly we label these four versions as ${\cal D}_{n}(\mu_{0})_{0}$, ${\cal D}_{n}^{d}(\mu_{0})_{0}$, ${\cal D}_{n}(\mu_{0})_{1}$ and ${\cal D}_{n}^{d}(\mu_{0})_{1}$. 

Table 4 presents the p-values associated with testing the global null of no predictability. We note strong rejections of the null across all implementations. US economic activity is clearly predictable using past macroeconomic information. More importantly our key-player estimator based on either of the four formulations of our test statistic identifies the same series given by the {\sf ISM: New Orders Index} (coded as NAPMNOI and numbered as {\sf 61} in FRED-MD). This is a monthly index published by the Institute for Supply Management and informing about the number of surveyed businesses reporting increased customer orders relative to the previous month. It is interesting that our key-player estimator has pointed to this predictor as NAPMNOI is the first forward looking indicator made public early each month. This variable has been found to be an important predictor of US recessions in Liu and Moench (2016) but perhaps more interestingly here, NAPMNOI can also be seen to be the most important predictor detected through the Bayesian methods developed in GLP2021. In their Figure 6 (GLP2021, p. 2425) this variable can be seen to be amongst the ones having the highest probabilities of inclusion.  The picture is even clearer in Fava and Lopes (2020, FL2020) who have reconsidered GLP2021's study by evaluating its sensitivity to the chosen priors. Figure 7 in FL2020 clearly points to the $61^{th}$ predictor as the one having a 100\% probability of of inclusion as a predictor of the US growth rate in industrial production.      
\vspace{-0.2cm}

\begin{table}[h]
	\centering
\caption{p-values} \vspace{0.5cm}
\scalebox{0.8}{\begin{tabular}{lcccclcccc} \hline
		$\mu_{0}$ & 0.30  & 0.35  & 0.40  & 0.45  &       & 0.30  & 0.35  & 0.40  & 0.45 \\
		${\cal D}_{n}(\mu_{0})_0$ & 0.000 & 0.001 & 0.003 & 0.001 & ${\cal D}_{n}(\mu_{0})_1$ & 0.000 & 0.001 & 0.002 & 0.000 \\
		${\cal D}_{n}^{d}(\mu_{0})_0$ & 0.000 & 0.000 & 0.000 & 0.000 & ${\cal D}_{n}^{d}(\mu_{0})_1$ & 0.000 & 0.000 & 0.000 & 0.000 \\
	\end{tabular}}
\label{tab:Tab5}
\end{table}

Although going beyond the detection of a key-player is outside the scope of this paper it is nevertheless interesting to evaluate leading predictors beyond the argmax of 
${\cal D}_{n}^{d}(\mu_{0})$. Table 5 isolates the top 6 predictors leading to the highest magnitudes of ${\cal D}_{n}^{d}(\mu_{0})$. We note a cluster of interest rate related predictors and a labour market indicator. It is again interesting to point out that predictors 92 and 39 also appear amongst the predictors with the highest probabilities of being included in both GLP2021 and FL2020. The heatmap presented in Figure 6 of GLP2021 (p. 2425) does also show a clustering of 
active predictors with IDs in the 90s range as in our Table 5. Lastly, it is also important to point out that the outcomes presented in Tables 4-5 remained unaffected when we have also augmented the pool of predictors to include the lagged growth rate in industrial production. This is in line with the fact that the growth rate in industrial production is only very weakly serially correlated.

\vspace{-0.28cm}
\begin{table}[h]
	\centering
	\caption{Key Player and Top 5 predictors} \vspace{0.5cm}
\scalebox{0.8}{\begin{tabular}{lll} \hline 
		Fred-MD ID & Fred-MD Code & Description \\ \hline
		61    & NAPMNOI & ISM: New Orders Index \\ \hline
		96    & T1YFFM & 1-Year Treasury C Minus FEDFUNDS \\
		95    & TB6SMFFM & 6-Month Treasury C Minus FEDFUNDS \\
		92    & BAA   & Moodys Seasoned Baa Corporate Bond Yield \\
		93    & COMPAPFFx & 3-Month Commercial Paper Minus FEDFUNDS \\
		39    & NDMANEMP & All Employees: Nondurable goods 
	\end{tabular}}
	\label{tab:Tab6}
\end{table}

\vspace{-1cm}
\section{Conclusions}

\vspace{-0.28cm}
We proposed a method for detecting the presence of out-of-sample predictability in the context of linear predictive regressions linking a response variable to one or more lagged predictors. 
An important novelty of our approach is its robustness to the dynamic properties of predictors which can be noisy, persistent or a mixture of both. In addition, our approach is able to accommodate a large number of predictors at little computational cost and has been shown to be very reliable even in contexts where the effective sample size is smaller than the available pool of predictors and this despite an asymptotic theory that operates solely under $n \rightarrow \infty$. As 
argued in McKeague and Qian (2008) who developed a theory of marginal screening in high dimensional regressions assuming a fixed pool of predictors, extending our own asymptotic framework that accommodates persistence and predictor correlatedness to environments in which $p$ is allowed to grow with $n$ would also raise formidable technical challenges. Nevertheless our local power analysis and simulation based results have clearly highlighted the suitability and accuracy of our asymptotic regime in large $p$ environments. 

In a wide range of applications one is often interested in whether a series is best described as a mean independent process or alternatively is characterized by predictability. This predictability may be driven by one or more predictors belonging to a very large information set.
One may not wish to take a stance on a particular predictor while also being constrained by dimensionality problems. 
The test we introduced in this paper is precisely designed to accommodate such 
environments. 
Although our primary focus is not about uncovering a true model our framework does allow us to detect a {\it key player} which can be valuable information in itself. It could for instance be used in a model further augmented with diffusion index type factors. It could also be useful in contexts where predictors consist solely of principal components as in such environments PC type factors are typically obtained while being agnostic about how they relate to the predictand. As PCs are typically obtained using a pool of very diverse predictors in terms of their persistence properties, the robustness of our methods to 
such characteristics also makes them particularly suitable. 
\vspace{-0.2cm}
\begin{center}
	{\bf REFERENCES}
\end{center}
\vspace{-0.2cm}

Berenguer-Rico, V. and B. Nielsen (2020). Cumulated Sum of Squares Statistics for Nonlinear and Nonstationary Regressions, \textit{Econometric Theory 36(1)}, 1-47. 




Clark, T.E. and K. D. West (2007). Approximately normal tests for equal predictive
accuracy in nested models, \textit{Journal of Econometrics 138 (1)}, 291-311. 



Clark, T.E. and M. W. McCracken (2013). Advances in Forecast Evaluation, in \textit{Handbook of Economic Forecasting},  Chapter 20, 1107-1201, Elsevier.

Corradi, V., Fosten J., and D. Gutknecht (2023). Predictive Ability Tests with Possibly Overlapping Models, SSRN Working Paper \href{http://dx.doi.org/10.2139/ssrn.4375650}{http://dx.doi.org/10.2139/ssrn.4375650.}

Decrouez, G. and P. Hall (2014). Split sample methods for constructing confidence intervals for binomial and Poisson parameters, \textit{J. R. Statist. Soc. B 76(5)}, 949-975. 

Deng, A. and P. Perron (2008). The Limit Distribution of the Cusum of Squares Test under General Mixing Conditions, \textit{Econometric Theory 24 (3)}, 809-822. 


Diebold, F.X. and R. Mariano (1995). Comparing Predictive Accuracy, \textit{Journal of Business and Economic Statistics 13 (3)}, 253-265. 

Fan, J. and J. Liv (2008). Sure independence screening for ultrah dimensional feature space, \textit{JRSS Series B, 70 (5)}, 849-911. 

Fan, J., Y. Liao, and J. Yao (2015). Power Enhancement in High Dimensional Cross-Sectional Tests, \textit{Econometrica 83 (4)}, 1497-1541. 

Fava, B. and H. F. Lopes (2020). The Illusion of the Illusion of Sparsity: An exercise in prior sensitivity, \textit{arXiv:2009.14296v1}, September 2020. 




Ghysels, E., J. B. Hill, and K. Moteigi (2020). Testing a large set of zero restrictions in regression models, with an application to mixed frequency Granger causality, \textit{Journal of
	Econometrics 218 (2)}, 633-654. 

Giannone, D., M. Lenza, and G. E. Primiceri (2021). Economic Predictions with Big Data: The Illusion of Sparsity, \textit{Econometrica}, 89, 2409-2437. 

Gonzalo, J. and J. Pitarakis (2019). Predictive Regressions, \textit{Oxford Research Encyclopedia: Economics and Finance}, November 2019. 

Goyal, A. and I. Welch (2008). A comprehensive Look at the Empirical Performance of Equity Premium Prediction, \textit{Review of Financial Studies, 21}, 1455-1508. 







Liu, W. and E. Moench (2016). What predicts US recessions?, \textit{International Journal of Forecasting}, 32, 1138-1150. 

McCracken, M. (2007). Asymptotics for out of sample tests of Granger causality, \textit{Journal of Econometrics 140 (2)}, 719-752.

McKeague, I. W. and M. Qian (2015). An adaptive resampling test for detecting the presence of significant predictors, \textit{Journal of the American Statistical Association 110 (512)}, 1442-1433. 

McCracken, M. and S. Ng (2016). FRED-MD: A Monthly Database for Macroeconomic Research, \textit{Journal of Business and Economic Statistics,} 34, 574-589. 




Pesaran, M. H. and A. Timmermann (1995). Predictability of Stock Returns: Robustness and Economic Significance,\textit{Journal of Finance}, 50, 1201-1228. 



Phillips, P. C. B. and A. Magdalinos (2009). Limit theory for cointegrated systems with moderately integrated and moderately explosive regressors, {\it Econometric Theory 25 (2)}, 482-526. 

Pitarakis, J. (2023). A Novel Approach to Predictive Accuracy Testing in Nested Environments, {\it Econometric Theory}, In Press.  


Schennah, S. M. and D. Wilhelm (2017). A Simple Parametric Model Selection Test, \textit{Journal of the American Statistical Association 112 (520)}, 1663-1674. 

Shi, X. (2015). A Nondegenerate Vuong Test, \textit{Quantitative Economics 6 (1)}, 85-121. 	


Stock, J. H. and Watson, M. W. (2002). Macroeconomic Forecasting Using Diffusion Indexes, \textit{Journal
	of Business and Economic Statistics 20}, 147-162.




Vuong, Q. H. (1989). Likelihood Ratio Tests for Model Selection and Non-nested Hypotheses, \textit{Econometrica 57 (2)}, 307-333. 	


West, K. (1996). Asymptotic Inference about Predictive Ability, \textit{Econometrica 64 (5)}, 1067-1084. 


West, K. (2006). Forecast Evaluation, in \textit{Handbook of Economic Forecasting}, Volume 1, 
Graham Elliott, Clive W.J. Granger and Allan Timmermann eds.




\newpage
\begin{center}
{\large\bf SUPPLEMENTARY MATERIAL}
\end{center}

\begin{description}
\item[] Technical Proofs and further simulations.
\end{description}

\def\spacingset#1{\renewcommand{\baselinestretch}%
	{#1}\small\normalsize} \spacingset{1}




\bigskip


\renewcommand{\thesection}{\Alph{section}}
\numberwithin{equation}{section}
\setcounter{section}{0}
\spacingset{1.8} 

This supplementary appendix contains two sections. Section A provides the technical proofs of the Propositions stated in the main text. Section B provides additional simulations that further illustrate the finite sample properties of our test and key player estimator. Emphasis is placed on various robustness considerations and a broader range of DGP parameterisations.

\section{PROOFS}
\label{sec:proofs}

\noindent
LEMMA A1: Let 
\begin{align}
	{\cal N}_{n}(m_{0}) &= \sqrt{n-k_{0}} \left(\dfrac{1}{2}\left( \dfrac{\sum_{t=k_{0}}^{k_{0}+m_{0}-1}\eta_{t+1}}{m_{0}}+\dfrac{\sum_{t=k_{0}+m_{0}}^{n-1}\eta_{t+1}}{n-k_{0}-m_{0}}\right)- \dfrac{\sum_{t=k_{0}}^{n-1}\eta_{t+1}}{n-k_{0}}\right) \label{eq:eq31}
\end{align}
with $\eta_{t}=(u_{t}^{2}-E[u_{t}^{2}])$ and $m_{0}=[(n-k_{0})\mu_{0}]$. The long run variance of 
${\cal N}_{n}(m_{0})$ is given by 
\begin{align}
	\omega^{2} &= \dfrac{(1-2\mu_{0})^{2}}{4\mu_{0}(1-\mu_{0})} \phi^{2} \label{eq:eq32}
\end{align}
\noindent where $\phi^{2}=\sum_{s=-\infty}^{\infty}\gamma_{\eta}(s)$. \\

\noindent
PROOF OF LEMMA A1. First, let us rewrite (\ref{eq:eq31}) as 
{\footnotesize 
	\begin{align}
		{\cal N}_{n}(m_{0}) &= \sqrt{n-k_{0}} \left(
		\dfrac{1}{2}\left(\dfrac{n-k_{0}}{m_{0}}\dfrac{\sum_{t=k_{0}}^{k_{0}+m_{0}-1} \eta_{t+1}}{n-k_{0}}+
		\dfrac{n-k_{0}}{n-k_{0}-m_{0}}\dfrac{\sum_{t=k_{0}+m_{0}}^{n-1} \eta_{t+1}}{n-k_{0}}\right)-
		\dfrac{\sum_{t=k_{0}}^{n-1} \eta_{t+1}}{n-k_{0}}
		\right) \label{eq:eq33}
	\end{align}
}
\noindent {\normalsize Using $m_{0}=[(n-k_{0})\mu_{0}]$ we have}
{\footnotesize 
	\begin{align}
		{\cal N}_{n}(\mu_{0}) &= 
		\sqrt{n-k_{0}} \left(
		\dfrac{1}{2}\left(\dfrac{1}{\mu_{0}} \dfrac{\sum_{t=k_{0}}^{k_{0}+m_{0}-1} \eta_{t+1}}{n-k_{0}}+
		\dfrac{1}{1-\mu_{0}} \dfrac{\sum_{t=k_{0}+m_{0}}^{n-1} \eta_{t+1}}{n-k_{0}}\right)-
		\dfrac{\sum_{t=k_{0}}^{n-1} \eta_{t+1}}{n-k_{0}}
		\right)+o(1). \label{eq:eq34}
	\end{align}
}
\normalsize
Next, let $I_{1t}\equiv I(k_{0}\leq t<k_{0}+m_{0})$ and $I_{2t}\equiv I(k_{0}+m_{0}\leq t<n)$ and define
\begin{align}
	Z_{t} &=  \eta_{t+1} \left(\dfrac{1}{2\mu_{0}} I_{1t}+\dfrac{1}{2(1-\mu_{0})} I_{2t}-1\right) \nonumber \\
	& \equiv \eta_{t+1} g_{t} \label{eq:eq35}
\end{align}
\noindent
so that (\ref{eq:eq34}) can be reformulated as
\begin{align}
	{\cal N}_{n}(\mu_{0}) &= \sqrt{n-k_{0}} \ \overline{Z}_{n}+o(1) \label{eq:eq36}
\end{align}
where $\overline{Z}_{n}=\sum_{t=k_{0}}^{n-1}Z_{t}/(n-k_{0})$. Standard algebra now leads to 
\begin{align}
	V\left[\sum_{t=k_{0}}^{n-1} Z_{t}\right] &= \sum_{t=k_{0}}^{n-1}V[Z_{t}]+ \sum_{t \neq s} Cov[Z_{t},Z_{s}] \nonumber \\
	&= (n-k_{0})\gamma_{Z}(0)+2 \sum_{s}(n-k_{0}-s)\gamma_{Z}(s) \label{eq:eq37}
\end{align}
\noindent where $\gamma_{Z}(0)=E[\eta_{t+1}^{2}g_{t}^{2}]$ and 
$\gamma_{Z}(s)=E[\eta_{t+1}\eta_{t+1-s}g_{t}g_{t-s}]$. Recalling the expression of the deterministic term $g_{t}$ from (\ref{eq:eq35}) it now suffices to note that $E[\eta_{t+1}^{2}g_{t}^{2}]=E[g_{t}^{2}]E[\eta_{t+1}^{2}]$ 
and $E[\eta_{t+1}\eta_{t+1-s}g_{t}g_{t-s}]=E[g_{t}^{2}]E[\eta_{t+1}\eta_{t+1-s}]$ with 
\begin{align}
	E[g_{t}^{2}] &= \dfrac{(1-2\mu_{0})^{2}}{4\mu_{0}(1-\mu_{0})} \label{eq:eq38}
\end{align}
from which it immediately follows that as $n \rightarrow \infty$
\begin{align}
	V[{\cal N}_{n}(\mu_{0})] & \rightarrow  \dfrac{(1-2\mu_{0})^{2}}{4\mu_{0}(1-\mu_{0})}\ \phi^{2} 	\label{eq:eq39}
\end{align}
for $\phi^{2}=\sum_{s=-\infty}^{\infty}\gamma_{\eta}(s)$. \hfill $\blacksquare$ \\

\noindent
PROOF OF PROPOSITION 1.  Under the null hypothesis Assumption 1(iii) implies that for given $\mu_{0}$ we 
can write 
{
	\begin{align}
		{\cal D}_{n}(m_{0},j) &= \dfrac{\sqrt{n-k_{0}}}{\hat{\omega}_{n}} \left(\dfrac{1}{2}\left( \dfrac{\sum_{t=k_{0}}^{k_{0}+m_{0}-1}(u_{t+1}^{2}-\sigma^{2}_{u})}{m_{0}}+\dfrac{\sum_{t=k_{0}+m_{0}}^{n-1}(u_{t+1}^{2}-\sigma^{2}_{u})}{n-k_{0}-m_{0}}\right)- \dfrac{\sum_{t=k_{0}}^{n-1}(u_{t+1}^{2}-\sigma^{2}_{u})}{n-k_{0}}\right) \nonumber \\
		&+ o_{p}(1). \label{eq:eq40}
	\end{align}
}
\noindent 
It now follows directly from Assumptions 1(i)-(ii) and the continuous mapping theorem that as $n \rightarrow \infty$
\begin{align}
	{\cal D}_{n}(\mu_{0},j) & \stackrel{d}\rightarrow  
	\dfrac{1}{\omega} \left(\dfrac{\phi}{2} 
	\left(\dfrac{W(\mu_{0})}{\mu_{0}}+\dfrac{W(1)-W(\mu_{0})}{1-\mu_{0}}
	\right)-\phi W(1)
	\right). \label{eq:eq41}
\end{align}

\noindent As we operate under a given $\mu_{0}$ it is now straightforward to observe that the variance of the expression between brackets in (\ref{eq:eq41}) is given by $\phi^{2}(1-2\mu_{0})^{2}/(4\mu_{0}(1-\mu_{0}))$. As 
$\hat{\omega} \stackrel{p}\rightarrow \omega=\phi (1-2\mu_{0})/\sqrt{4 \mu_{0}(1-\mu_{0})}$
it follows from Slutsky's theorem that ${\cal D}_{n}(\mu_{0},j) \stackrel{d}\rightarrow {\cal Z}$. 
Recalling that we operate under a fixed number of predictors $p$ and as the distributional limit in (2) holds for all $j \in \{1,2,\ldots,p\}$ we have from (1) that 
$\sum_{j=1}^{p}\lambda_{j}{\cal D}_{n}(m_{0},j) \stackrel{d}\rightarrow (\sum_{j=1}^{p}\lambda_{j}) {\cal Z}$ so that joint convergence follows from the Cramer-Wold device with $({\cal D}_{n}(m_{0},j),\ldots,{\cal D}_{n}(m_{0},p))\stackrel{d}\rightarrow ({\cal Z},\ldots,{\cal Z})$ and the stated result for $\overline{\cal D}_{n}(\mu_{0})\stackrel{d}\rightarrow {\cal Z}$ follows.   
\hfill $\blacksquare$ \\

Before proceeding with the proofs of Propositions 2A, 2B and 2C we introduce a series of intermediate results and further notation that will be used throughout. As we operate under the hypothesis of at least one active predictor the true {\it local to the null} specifications under the three scenarios A, B and C are understood to be given by 
\begin{align}
	y_{t+1} &= \sum_{i \in {\cal I}^{*}} (\beta_{i}^{*}/n^{1/4}) \ x_{it}+ u_{t+1} \label{eq:eq42}
\end{align}
\begin{align}
	y_{t+1} &= \sum_{i \in {\cal I}^{*}} (\beta_{i}^{*}/n^{(1+2\alpha)/4}) \ x_{it}+ u_{t+1} \label{eq:eq43}
\end{align}
and 
\begin{align}
	y_{t+1} &= \sum_{i \in {\cal I}^{*}_{1}} (\beta_{i}^{*}/n^{1/4}) \ x_{it}+\sum_{i \in {\cal I}^{*}_{2}} (\beta_{i}^{*}/n^{(1+2\alpha)/4}) \ x_{it}+ u_{t+1} \label{eq:eq44}
\end{align}
\noindent respectively. We also recall that the fitted specification involving one predictor at a time is 
here given by 
\begin{align}
	y_{t+1} &= \beta_{j}x_{jt}+u_{t+1} \ \ \ \ j=1,\ldots,p \label{eq:eq45}
\end{align}
\noindent so that using (\ref{eq:eq42}) and (\ref{eq:eq43}) we can write the recursively estimated slope parameters as
\begin{align}
	\hat{\beta}_{jt} &= \dfrac{\sum_{i \in {\cal I}^{*}} \beta_{i}^{*}(\sum_{s=1}^{t}x_{is}x_{js})}{n^{\gamma}\sum_{s=1}^{t}x_{js}^{2}}+\dfrac{\sum_{s=1}^{t}x_{js}u_{s+1}}{\sum_{s=1}^{t}x_{js}^{2}} \label{eq:eq46}
\end{align}
where $\gamma=1/4$ under scenario A and $\gamma=(1+2\alpha)/4$ under scenario B. For the mixed predictor scenario C and using (\ref{eq:eq44}) we have instead
\begin{align}
	\hat{\beta}_{jt} &=  \dfrac{\sum_{i \in {\cal I}^{*}_{1}} \beta_{i}^{*}(\sum_{s=1}^{t}x_{is}x_{js})}{n^{\gamma_{1}}\sum_{s=1}^{t}x_{js}^{2}}+\dfrac{\sum_{i \in {\cal I}^{*}_{2}} \beta_{i}^{*}(\sum_{s=1}^{t}x_{is}x_{js})}{n^{\gamma_{2}}\sum_{s=1}^{t}x_{js}^{2}}+\dfrac{\sum_{s=1}^{t}x_{js}u_{s+1}}{\sum_{s=1}^{t}x_{js}^{2}}. \label{eq:eq47}
\end{align}

\noindent
The specifications in (\ref{eq:eq42})-(\ref{eq:eq44}) are the DGPs under the local 
alternatives of interest and the $\hat{\beta}_{jt}$'s in (\ref{eq:eq46})-(\ref{eq:eq47}) are the slope parameters estimated via recursive least squares when fitting (\ref{eq:eq45}). As for notational convenience we have abstracted from the inclusion of an intercept in the above specifications it is naturally understood that the forecasts under the null model will be taken as $\hat{y}_{0,t+1|t}=0$ instead of 
$\sum_{j=1}^{t}y_{j}/t$. This has no bearing on any of the asymptotic results presented in Propositions 2A-2C. We can now write the forecast errors as
\begin{align}
	\hat{e}_{0,t+1|t} &= y_{t+1}-0 \nonumber \\
	\hat{e}_{j,t+1|t} &= y_{t+1}-\hat{\beta}_{jt} x_{jt} \label{eq:eq48}
\end{align}
\noindent with $y_{t+1}$ given by either (\ref{eq:eq42}), (\ref{eq:eq43}) or (\ref{eq:eq44}). \\

\noindent
LEMMA A2. Under Assumption 2A, $\hat{\beta}_{jt}$ as in (\ref{eq:eq46}) and $\forall j \in \{1,\ldots,p\}$ we have  as $n\rightarrow \infty$
\begin{enumerate}
	\item[(i)] $\sup_{r \in [\pi_{0},1]}\left|n^{1/4}\hat{\beta}_{j,[nr]}-\dfrac{1}{E[x_{jt}^{2}]}
	\sum_{i\in {\cal I}^{*}} \beta_{i}^{*} E[x_{it}x_{jt}] \right|=o_{p}(1)$
	\item[(ii)] $\sup_{k_{0}\leq t\leq n }\left|
	\dfrac{\sum_{\ell=k_{0}}^{t}\hat{\beta}_{j\ell}x_{j\ell}u_{\ell+1}}{\sqrt{n-k_{0}}}
	\right|=o_{p}(1)$
	\item[(iii)] $\sup_{k_{0}\leq t\leq n }\left|
	\dfrac{\sum_{\ell=k_{0}}^{t}\hat{\beta}_{j\ell}^{2}x_{j\ell}^{2}}{\sqrt{n-k_{0}}}- \dfrac{\sqrt{1-\pi_{0}}}{E[x_{jt}^{2}]}
	\left(\sum_{i\in {\cal I}^{*}} \beta_{i}^{*} 
	E[x_{it}x_{jt}]\right)^{2}
	\right|=o_{p}(1)$
	\item[(iv)] $\sup_{k_{0}\leq t\leq n }\left|\dfrac{\beta_{i}^{*}}{n^{1/4}}
	\dfrac{\sum_{\ell=k_{0}}^{t}\hat{\beta}_{j\ell}x_{i\ell}x_{j\ell}}{\sqrt{n-k_{0}}}-\sqrt{1-\pi_{0}} \
	\beta_{i}^{*}\dfrac{E[x_{it}x_{jt}]}{E[x_{jt}^{2}]}
	(\sum_{i \in {\cal I}^{*}}\beta_{i}^{*}E[x_{it}x_{jt}])
	\right|=o_{p}(1)$	
\end{enumerate}
\vspace{0.4cm}

\noindent PROOF OF LEMMA A2. (i) From (\ref{eq:eq46}) we have 
\begin{align}
	n^{1/4} \hat{\beta}_{jt} &=  \dfrac{\sum_{i \in {\cal I}^{*}} \beta_{i}^{*}(\sum_{s=1}^{t}x_{is}x_{js})}{\sum_{s=1}^{t}x_{js}^{2}}+n^{1/4}\dfrac{\sum_{s=1}^{t}x_{js}u_{s+1}}{\sum_{s=1}^{t}x_{js}^{2}} \label{eq:eq49}
\end{align}
\noindent and 
\begin{align}
	n^{1/4} \sup_{t}|\hat{\beta}_{jt}| & \leq \sup_{t}\left|\dfrac{\sum_{i \in {\cal I}^{*}} \beta_{i}^{*}(\sum_{s=1}^{t}x_{is}x_{js})}{\sum_{s=1}^{t}x_{js}^{2}}\right|+n^{1/4}\sup_{t}\left|\dfrac{\sum_{s=1}^{t}x_{js}u_{s+1}}{\sum_{s=1}^{t}x_{js}^{2}}\right|. \label{eq:eq50}
\end{align}
\noindent
We can now note that 
\begin{align}
	n^{1/4} \sup_{t}\left|\dfrac{\sum_{s=1}^{t}x_{js}u_{s+1}}{\sum_{s=1}^{t}x_{js}^{2}}\right| & \leq \sup_{t}\left|\dfrac{t}{\sum_{s=1}^{t} x_{js}^{2}}\right|\dfrac{n^{1/4}}{t}\sup_{t} \left|\sum_{s=1}^{t}x_{js}u_{s+1}\right|\stackrel{p}\rightarrow 0 \label{eq:eq51}
\end{align}
\noindent which follows directly from Assumption 2A(iii). This latter assumption now also leads to 
\begin{align}
	\sup_{t} \left|\dfrac{\sum_{i \in {\cal I}^{*}} \beta_{i}^{*}(\sum_{s=1}^{t}x_{is}x_{js})}{\sum_{s=1}^{t}x_{js}^{2}}-\sum_{i\in {\cal I}^{*}}\beta_{i}^{*}\dfrac{E[x_{it}x_{jt}]}{E[x_{jt}^{2}]}\right| &= o_{p}(1)  \label{eq:eq52}
\end{align}

\noindent as required. (ii) We write 
\begin{align}
	\sup_{k_{0}\leq t\leq n }\left|
	\dfrac{\sum_{\ell=k_{0}}^{t}\hat{\beta}_{j\ell}x_{j\ell}u_{\ell+1}}{\sqrt{n-k_{0}}}
	\right| &= \dfrac{1}{\sqrt{1-\pi_{0}}}\dfrac{1}{n^{1/4}} \sup_{r \in [\pi_{0},1]} \left|
	\dfrac{\sum_{l=k_{0}}^{[nr]} (n^{1/4}\hat{\beta}_{j\ell}) x_{j\ell}u_{\ell+1}}{\sqrt{n}}
	\right|+o_{p}(1). \label{eq:eq53}
\end{align}
The result in part (i) combined with Assumption 2A(iii) allows us to appeal to Theorem 3.3 in Hansen (1993) from which the statement in (ii) follows. For part (iii) it is sufficient to focus on 
\begin{align}
	\dfrac{1}{\sqrt{n-k_{0}}}\sum_{\ell=k_{0}}^{n-1}\hat{\beta}_{j\ell}^{2}x_{j\ell}^{2} &= 
	\dfrac{1}{\sqrt{1-\pi_{0}}}\dfrac{1}{n}\sum_{\ell=k_{0}}^{n} (\sqrt{n}\hat{\beta}_{j\ell}^{2})x_{j\ell}^{2}+o(1)  \label{eq:eq54}
\end{align}
\noindent for which part (i) combined with Assumptions 2A(iii) ensures that 
\begin{align}
	\dfrac{1}{\sqrt{n-k_{0}}}\sum_{\ell=k_{0}}^{n-1}\hat{\beta}_{j\ell}^{2}x_{j\ell}^{2} & \stackrel{p}\rightarrow  \sqrt{1-\pi_{0}} \left(\sum_{i \in {\cal I}^{*}} \beta_{i}^{*}\dfrac{E[x_{it}x_{jt}]}{\sqrt{E[x_{jt}^{2}]}}\right)^{2}. \label{eq:eq55}
\end{align}
\noindent Part (iv) follows identical lines to part (iii) and its details are therefore omitted. 
\hfill $\blacksquare$
\vspace{0.6cm}

\noindent
PROOF OF PROPOSITION 2A. Using $y_{t+1}$ as in (\ref{eq:eq42}) in $\hat{e}_{0,t+1|t}^{2}=y_{t+1}^{2}$ from (\ref{eq:eq48}), we have

\begin{align}
	\dfrac{\sum_{t=k_{0}}^{k_{0}+m_{0}-1} \hat{e}_{0,t+1|t}^{2}}{\sqrt{n-k_{0}}} & = 
	\dfrac{\sum_{t=k_{0}}^{k_{0}+m_{0}-1} u_{t+1}^{2}}{\sqrt{n-k_{0}}}+
	\dfrac{\sum_{t=k_{0}}^{k_{0}+m_{0}-1} (\sum_{i \in {\cal I}^{*}}\beta_{i}^{*}x_{it})^{2}}{\sqrt{n}\sqrt{n-k_{0}}} \nonumber \\ 
	& + \dfrac{2}{n^{1/4}} \sum_{i\in {\cal I}^{*}} \beta_{i}^{*}\left(\dfrac{\sum_{t=k_{0}}^{k_{0}+m_{0}-1}x_{it}u_{t+1}}{\sqrt{n-k_{0}}} \right) \nonumber \\
	& = 
	\dfrac{\sum_{t=k_{0}}^{k_{0}+m_{0}-1} u_{t+1}^{2}}{\sqrt{n-k_{0}}}+
	\dfrac{\sum_{t=k_{0}}^{k_{0}+m_{0}-1} (\sum_{i \in {\cal I}^{*}}\beta_{i}^{*}x_{it})^{2}}{\sqrt{n}\sqrt{n-k_{0}}} +o_{p}(1) \nonumber \\
	& = \dfrac{\sum_{t=k_{0}}^{k_{0}+m_{0}-1} u_{t+1}^{2}}{\sqrt{n-k_{0}}}+\mu_{0}\sqrt{1-\pi_{0}} \
	E\left[
	\sum_{i\in {\cal I}^{*}}\beta_{i}^{*}x_{it}
	\right]^{2}+ o_{p}(1) \label{eq:eq56}
\end{align}
\noindent where we made repeated use of Assumption 2A(iii). Proceeding as above it also follows that 
\begin{align}
	\dfrac{\sum_{t=k_{0}+m_{0}}^{n-1} \hat{e}_{0,t+1|t}^{2}}{\sqrt{n-k_{0}}} & =  \dfrac{\sum_{t=k_{0}+m_{0}}^{n-1} u_{t+1}^{2}}{\sqrt{n-k_{0}}}+(1-\mu_{0})\sqrt{1-\pi_{0}} \
	E\left[
	\sum_{i\in {\cal I}^{*}}\beta_{i}^{*}x_{it}
	\right]^{2}+ o_{p}(1). \label{eq:eq57}
\end{align}
\noindent Next, we focus on $\hat{e}_{j,t+1|t}^{2}$ given by (\ref{eq:eq48}) with $y_{t+1}$ as in (\ref{eq:eq42}). We have 

\begin{align}
	\dfrac{\sum_{t=k_{0}}^{n-1} \hat{e}_{j,t+1|t}^{2}}{\sqrt{n-k_{0}}} & =  
	\dfrac{\sum_{t=k_{0}}^{n-1} u_{t+1}^{2}}{\sqrt{n-k_{0}}}+
	\dfrac{1}{\sqrt{n(n-k_{0})}} \sum_{t=k_{0}}^{n-1}(\sum_{i \in {\cal I}^{*}} \beta_{i}^{*}x_{it})^{2} \nonumber \\ & + 
	\dfrac{2}{n^{1/4}\sqrt{n-k_{0}}} 
	\sum_{i \in {\cal I}^{*}}\beta_{i}^{*} 
	(\sum_{t=k_{0}}^{n-1} x_{it}u_{t+1})+
	\dfrac{1}{\sqrt{n-k_{0}}}\sum_{t=k_{0}}^{n-1}\hat{\beta}_{jt}^{2}x_{jt}^{2} \nonumber \\ & -  \dfrac{2}{n^{1/4}\sqrt{n-k_{0}}} \sum_{i\in {\cal I}^{*}} \beta_{i}^{*}(\sum_{t=k_{0}}^{n-1} \hat{\beta}_{jt}x_{jt}x_{it}) \nonumber \\
	& -  \dfrac{2}{\sqrt{n-k_{0}}} \sum_{t=k_{0}}^{n-1}\hat{\beta}_{jt}x_{jt}u_{t+1}. \label{eq:eq58}
\end{align}
Appealing to Assumption 2A(iii) and using Lemma A2(ii)-(iii) in (\ref{eq:eq58}) also allows us to write
\begin{align}
	\dfrac{\sum_{t=k_{0}}^{n-1} \hat{e}_{j,t+1|t}^{2}}{\sqrt{n-k_{0}}} & = 
	\dfrac{\sum_{t=k_{0}}^{n-1} u_{t+1}^{2}}{\sqrt{n-k_{0}}}+
	\sqrt{1-\pi_{0}} \ E\left[\sum_{i \in {\cal I}^{*}} \beta_{i}^{*}x_{it}\right]^{2} \nonumber \\ & - 
	\dfrac{\sqrt{1-\pi_{0}}}{E[x_{jt}^{2}]} 
	\left(
	\sum_{i \in {\cal I}^{*}} \beta_{i}^{*}E[x_{it}x_{jt}]
	\right)^{2}
	+o_{p}(1). \label{eq:eq59}
\end{align}
\noindent 
Next, using (\ref{eq:eq56})-(\ref{eq:eq59}) in ${\cal D}_{n}(m_{0},j)$ now gives
\begin{align}
	{\cal D}_{n}(m_{0},j) & = \dfrac{1}{\omega(\mu_{0})} \left(\dfrac{1}{2}\left(\dfrac{n-k_{0}}{m_{0}} \dfrac{\sum_{t=k_{0}}^{k_{0}+m_{0}-1} u_{t+1}^{2}}{\sqrt{n-k_{0}}}+\dfrac{n-k_{0}}{n-k_{0}-m_{0}}
	\dfrac{\sum_{t=k_{0}+m_{0}}^{n-1} u_{t+1}^{2}}{\sqrt{n-k_{0}}}\right)-\dfrac{\sum_{t=k_{0}}^{n-1} u_{t+1}^{2}}{\sqrt{n-k_{0}}}\right) \nonumber \\
	& + \sqrt{1-\pi_{0}} \dfrac{1}{\omega(\mu_{0})}
	\left(
	\sum_{i \in {\cal I}^{*}} \beta_{i}^{*}
	\dfrac{E[x_{it}x_{jt}]}{\sqrt{E[x_{jt}^{2}]}}\right)^{2}+o_{p}(1) \label{eq:eq60}
\end{align}
\noindent with $\omega(\mu_{0})$ as in (6) and thus leading to the desired result in (14) of Proposition 2A. \hfill $\blacksquare$ \\

\noindent 
LEMMA B1. Under Assumption 2B, $\hat{\beta}_{jt}$ as in (\ref{eq:eq46}) and $\forall j \in \{1,\ldots,p\}$ we have as $n\rightarrow \infty$
\begin{enumerate}
	\item[(i)] $\sup_{r \in [\pi_{0},1]}\left|n^{(1+2\alpha)/4}\hat{\beta}_{j,[nr]}-
	\sum_{i\in {\cal I}^{*}} \beta_{i}^{*} \  \dfrac{\sigma_{v_{i}v_{j}}}{\sigma^{2}_{v_{j}}} \left(\dfrac{2c_{j}}{c_{i}+c_{j}}\right) \right|=o_{p}(1)$
	\item[(ii)] $\sup_{k_{0}\leq t\leq n }\left|
	\dfrac{\sum_{\ell=k_{0}}^{t}\hat{\beta}_{j\ell}x_{j\ell}u_{\ell+1}}{\sqrt{n-k_{0}}}
	\right|=o_{p}(1)$
	\item[(iii)] $\sup_{k_{0}\leq t\leq n }\left|
	\dfrac{\sum_{\ell=k_{0}}^{t}\hat{\beta}_{j\ell}^{2}x_{j\ell}^{2}}{\sqrt{n-k_{0}}}-\sqrt{1-\pi_{0}}
	\left(\sum_{i \in {\cal I}^{*}}\beta_{i}^{*} \dfrac{\sigma_{v_{i}v_{j}}}{\sqrt{\sigma^{2}_{v_{j}}}} 
	\dfrac{\sqrt{2 c_{j}}}{c_{i}+c_{j}}\right)^{2}
	\right|=o_{p}(1)$,
	\item[(iv)] $\sup_{k_{0}\leq t\leq n }\left|\dfrac{\beta_{i}^{*}}{n^{(1+2\alpha)/4}}
	\dfrac{\sum_{\ell=k_{0}}^{t}\hat{\beta}_{j\ell}x_{i\ell}x_{j\ell}}{\sqrt{n-k_{0}}}-\sqrt{1-\pi_{0}} 
	\dfrac{2 c_{j}\sigma_{v_{i}v_{j}}}{(c_{i}+c_{j})\sigma^{2}_{v_{j}}}
	\left(\sum_{i \in {\cal I}^{*}} \beta_{i}^{*} \dfrac{\sigma_{v_{i}v_{j}}}{c_{i}+c_{j}}\right)
	\right|=o_{p}(1)$	
\end{enumerate}

\vspace{0.2cm}
\noindent
PROOF OF LEMMA B1. For all four cases the results follow in an identical manner to Lemma A2(i)-(iv) with the use of Assumption 2A(iii) replaced with Assumption 2B(iii) and $n^{1/4}$ replaced with $n^{(1+2\alpha)/4}$. \hfill $\blacksquare$ \\

\noindent 
PROOF OF PROPOSITION 2B. Using $y_{t+1}$ as in (\ref{eq:eq43}) in $\hat{e}_{0,t+1|t}=y_{t+1}^{2}$ from (\ref{eq:eq48}) we have
\begin{align}
	\dfrac{\sum_{t=k_{0}}^{k_{0}+m_{0}-1} \hat{e}_{0,t+1|t}^{2}}{\sqrt{n-k_{0}}} & =  
	\dfrac{\sum_{t=k_{0}}^{k_{0}+m_{0}-1} u_{t+1}^{2}}{\sqrt{n-k_{0}}}+
	\dfrac{\sum_{t=k_{0}}^{k_{0}+m_{0}-1} (\sum_{i \in {\cal I}^{*}}\beta_{i}^{*}x_{it})^{2}}{n^{(1+2\alpha)/2}\sqrt{n-k_{0}}} \nonumber \\ 
	& +  \dfrac{2}{n^{(1+2\alpha)/4}} \sum_{i\in {\cal I}^{*}} \beta_{i}^{*}\left(\dfrac{\sum_{t=k_{0}}^{k_{0}+m_{0}-1}x_{it}u_{t+1}}{\sqrt{n-k_{0}}} \right) \nonumber \\
	& =  
	\dfrac{\sum_{t=k_{0}}^{k_{0}+m_{0}-1} u_{t+1}^{2}}{\sqrt{n-k_{0}}}+
	\dfrac{\sum_{t=k_{0}}^{k_{0}+m_{0}-1} (\sum_{i \in {\cal I}^{*}}\beta_{i}^{*}x_{it})^{2}}{{n^{(1+2\alpha)/2}\sqrt{n-k_{0}}}} +o_{p}(1) \label{eq:eq61}
\end{align}
\noindent
and
\begin{align}
	\dfrac{\sum_{t=k_{0}+m_{0}}^{n-1} \hat{e}_{0,t+1|t}^{2}}{\sqrt{n-k_{0}}} & =  
	\dfrac{\sum_{t=k_{0}+m_{0}}^{n-1} u_{t+1}^{2}}{\sqrt{n-k_{0}}}+
	\dfrac{\sum_{t=k_{0}+m_{0}}^{n-1} (\sum_{i \in {\cal I}^{*}}\beta_{i}^{*}x_{it})^{2}}{n^{(1+2\alpha)/2}\sqrt{n-k_{0}}} \nonumber \\ 
	& +  \dfrac{2}{n^{(1+2\alpha)/4}} \sum_{i\in {\cal I}^{*}} \beta_{i}^{*}\left(\dfrac{\sum_{t=k_{0}+m_{0}}^{n-1}x_{it}u_{t+1}}{\sqrt{n-k_{0}}} \right) \nonumber \\
	& =  
	\dfrac{\sum_{t=k_{0}+m_{0}}^{n-1} u_{t+1}^{2}}{\sqrt{n-k_{0}}}+
	\dfrac{\sum_{t=k_{0}+m_{0}}^{n-1} (\sum_{i \in {\cal I}^{*}}\beta_{i}^{*}x_{it})^{2}}{{n^{(1+2\alpha)/2}\sqrt{n-k_{0}}}} +o_{p}(1).  \label{eq:eq62}
\end{align}
\noindent Next, for $\hat{e}_{j,t+1|t}^{2}$ we have
\begin{align}
	\dfrac{\sum_{t=k_{0}}^{n-1} \hat{e}_{j,t+1|t}^{2}}{\sqrt{n-k_{0}}} & =  
	\dfrac{\sum_{t=k_{0}}^{n-1} u_{t+1}^{2}}{\sqrt{n-k_{0}}}+
	\dfrac{1}{n^{(1+2\alpha)/2}\sqrt{(n-k_{0})}} \sum_{t=k_{0}}^{n-1}(\sum_{i \in {\cal I}^{*}} \beta_{i}^{*}x_{it})^{2} \nonumber \\ & -
	\sqrt{1-\pi_{0}} \ \dfrac{2c_{j}}{\sigma^{2}_{v_{j}}} \left(
	\sum_{i \in {\cal I}^{*}}\beta_{i}^{*} \dfrac{\sigma_{v_{i}v_{j}}}{c_{i}+c_{j}}\right)^{2}+o_{p}(1).  \label{eq:eq63}
\end{align}
\noindent Using (\ref{eq:eq61})-(\ref{eq:eq63}) in ${\cal D}_{n}(m_{0},j)$ and rearranging gives
\begin{align}
	{\cal D}_{n}(m_{0},j) & = \dfrac{1}{\omega(\mu_{0})} \left(\dfrac{1}{2}\left(\dfrac{n-k_{0}}{m_{0}} \dfrac{\sum_{t=k_{0}}^{k_{0}+m_{0}-1} u_{t+1}^{2}}{\sqrt{n-k_{0}}}+\dfrac{n-k_{0}}{n-k_{0}-m_{0}}
	\dfrac{\sum_{t=k_{0}+m_{0}}^{n-1} u_{t+1}^{2}}{\sqrt{n-k_{0}}}\right)- \dfrac{\sum_{t=k_{0}}^{n-1} u_{t+1}^{2}}{\sqrt{n-k_{0}}}\right) \nonumber \\
	& +  \sqrt{1-\pi_{0}} \dfrac{1}{\omega(\mu_{0})}
	\left(
	\sum_{i \in {\cal I}^{*}} \beta_{i}^{*}
	\dfrac{\sigma_{v_{i}v_{j}}}{\sqrt{\sigma^{2}_{v_{j}}}}
	\sqrt{\dfrac{2 c_{j}}{(c_{i}+c_{j})^{2}}}
	\right)^{2}+o_{p}(1)  \label{eq:eq64}
\end{align}
\noindent leading to the result in (18) of Proposition 2B. \hfill $\blacksquare$ \\

\noindent 
LEMMA C1. Under Assumption 2C, $\hat{\beta}_{jt}$ as in (\ref{eq:eq47}) and $\forall j \in \{1,\ldots,p\}$ we have as $n\rightarrow \infty$
\begin{enumerate}
	
	\item[(i)] $\sup_{r \in [\pi_{0},1]}\left|n^{1/4}\hat{\beta}_{j,[nr]}-\dfrac{1}{E[x_{jt}^{2}]}
	\sum_{i\in {\cal I}_{1}^{*}} \beta_{i}^{*} E[x_{it}x_{jt}] \right|=o_{p}(1)$ \ \ for $j \in {\cal I}_{1}^{*}$
	
	\item[(ii)] $\sup_{r \in [\pi_{0},1]}\left|n^{(1+2\alpha)/4}\hat{\beta}_{j,[nr]}-
	\sum_{i\in {\cal I}_{2}^{*}} \beta_{i}^{*} \  \dfrac{\sigma_{v_{i}v_{j}}}{\sigma^{2}_{v_{j}}} \left(\dfrac{2c_{j}}{c_{i}+c_{j}}\right) \right|=o_{p}(1)$ \ \ for $j \in {\cal I}_{2}^{*}$.
\end{enumerate}

\vspace{0.2cm}
\noindent
PROOF OF LEMMA C1. (i) and (ii) are obtained following the same derivations as Lemma A2(i) and LemmaA B1(i) and details are therefore omitted. 
\hfill $\blacksquare$ \\

\noindent 
PROOF OF PROPOSITION 2C. The result in (21) is obtained following identical derivations to (14) and
(18) and details are therefore omitted. \hfill $\blacksquare$ \\


\vspace{0.2cm}
\noindent
PROOF OF PROPOSITION 3. (i) It suffices to establish that under the null hypothesis $\widetilde{d}_{nj}=o_{p}(1)$. We operate under the benchmark model in (2) and with no loss of generality set the intercept to zero. The null model is given by $y_{t+1}=u_{t+1}$ so that $\hat{e}_{0,t+1|t}=u_{t+1}$ and 
$\hat{e}_{j,t+1|t}=u_{t+1}-\hat{\beta}_{jt}x_{jt}$
for $\hat{\beta}_{jt}=\sum_{s=1}^{t}x_{js}u_{s+1}/\sum_{s=1}^{t}x_{js}^{2}$.
It now follows that $d_{nj}=(n-k_{0})^{-1}\sum_{t=k_{0}}^{n-1} \hat{\beta}_{jt}^{2}x_{jt}^{2}$. We can now use Lemma A2(ii) specialised to $\beta_{i}^{*}=0$ $\forall i$ to infer that $d_{nj}=(n-k_{0})^{-1/2} o_{p}(1)$ 
so that $\widetilde{d}_{nj}=o_{p}(1)$. This ensures that our augmented test statistic maintains the same limiting distribution as ${\cal D}_{n}(m_{0})$. (ii) Here we treat the case of Scenario A as the remaining cases follow exactly the same line of proof. For scenario A the true model is given by (\ref{eq:eq42}) and the quantity of interest is given by 
$\widetilde{d}_{nj}=\hat{\omega}_{n}^{-1} (n-k_{0})^{-1/2} \sum_{t=k_{0}}^{n-1}(\hat{e}_{0,t+1|t}-\hat{e}_{j,t+1|t})^{2}$ from which 
we have $\widetilde{d}_{nj}=\hat{\omega}_{n}^{-1} (n-k_{0})^{-1/2} \sum_{t=k_{0}}^{n-1}\hat{\beta}_{jt}^{2}x_{jt}^{2}$ for 
$\hat{\beta}_{jt}=\sum_{s=1}^{t}x_{js}y_{s+1}/\sum_{s=1}^{t}x_{js}^{2}$. Appealing to Lemma A2(ii) it now immediately follows that 
$\widetilde{d}_{nj}\stackrel{d}\rightarrow g(\mu_{0},\pi_{0},\phi) 
\sum_{j=1}^{p}(\sum_{i \in \cal I^{*}} \beta_{i}^{*} E[x_{it}x_{jt}]/\sqrt{E[x_{jt}^{2}]})^{2}/p$. The result in Proposition 3(ii) now follows directly from the definition of $\overline{\cal D}_{n}(m_{0})$ in (22). \\

\noindent 
PROOF OF PROPOSITION 4. For part (i) of Proposition 4 we focus solely on  the case of a single stationary active predictor in the DGP as the remaining scenarios follow identical lines. It is useful to first note that the argmax of ${\cal D}_{n}(m_{0},j)$ will be equivalent to $\arg \min_{j} {\cal S}_{n}(j)$ where  
\begin{align}
	{\cal S}_{n}(j) & = \dfrac{\sum_{t=k_{0}}^{n-1}(\hat{e}_{j,t+1|t}^{2}-u_{t+1}^{2})}{\sqrt{n-k_{0}}} \ \ \ j=1,\ldots,p.  \label{eq:eq65}
\end{align}
\noindent The main result now follows by establishing that ${\cal S}_{n}(j)$ converges to a deterministic limit that is 
uniquely minimized at $j=j_{0}$. We continue to operate under the DGP given by (\ref{eq:eq42}) with $|{\cal I}^{*}|=1$ (i.e. there is a single active predictor) and with no loss of generality we set that predictor to be $x_{1t}$. 
Recalling that $\hat{e}_{j,t+1|t}=y_{t+1}-\hat{\beta}_{jt}x_{jt}$ and using Lemma A2 it immediately follows that for $j=j_{0}=1$ we have ${\cal S}_{n}(j=1)\stackrel{p}\rightarrow 0$ while for $j \neq j_{0}=1$ and using Lemma A2
we have 
\begin{align}
	{\cal S}_{n}(j) & \stackrel{p}\rightarrow (\beta_{1}^{*})^{2}\sqrt{1-\pi_{0}} E[x_{1t}^{2}](1-\rho^{2}_{1j}) \ \ \ \forall j\neq j_{0}  \label{eq:eq66}
\end{align} 
\noindent which is strictly positive for any predictor different from $x_{1t}$, thus leading to the required result. (ii) For part (ii) of Proposition 4 we consider the DGP given by (\ref{eq:eq44}) and that consists of predictors with mixed persistence properties. We operate with a pool of $p_{1}$ stationary predictors and $p-p_{1}\equiv p_{2}$ persistent predictors and with no loss of generality take $j=1,\ldots,p_{1}$ to index the stationary predictors and $j=p_{1}+1,\ldots,p$ the persistent predictors. We assume two active predictors given by $x_{at}=x_{1t}$ and $x_{bt}=x_{p_{1}+1,t}$ respectively. Using the results in Lemmas A1, B1 and C1 and standard algebra gives
\begin{align}
	{\cal S}_{n}(j=1) & \stackrel{p}\rightarrow  \sqrt{1-\pi_{0}} (\beta_{p_{1}+1}^{*})^{2}\dfrac{\sigma^{2}_{v_{p_{1}+1}}}{2 c_{p_{1}+1}}  \label{eq:eq67} \\
	{\cal S}_{n}(j \in \{2,\ldots,p_{1}\}) & \stackrel{p}\rightarrow  \sqrt{1-\pi_{0}} (\beta_{p_{1}+1}^{*})^{2}\dfrac{\sigma^{2}_{v_{p_{1}+1}}}{2 c_{p_{1}+1}}+ \sqrt{1-\pi_{0}} (\beta_{1}^{*})^{2}E[x_{1t}^{2}](1-\rho^{2}_{1j})   \label{eq:eq68} \\
	{\cal S}_{n}(j=p_{1}+1) & \stackrel{p}\rightarrow  \sqrt{1-\pi_{0}} (\beta_{1}^{*})^{2}E[x_{1t}^{2}]   \label{eq:eq69} \\
	{\cal S}_{n}(j\in \{p_{1}+2,\ldots,p\}) & \stackrel{p}\rightarrow 
	\sqrt{1-\pi_{0}}(\beta_{1}^{*})^{2}E[x_{1t}^{2}]+\sqrt{1-\pi_{0}}(\beta_{p_{1}+1}^{*})^{2} \dfrac{\sigma^{2}_{v_{p_{1}+1}}}{2 c_{p_{1}+1}} \nonumber \\ 
	& \times  \left(1-\dfrac{(\sigma_{p_{1}+1,j}/(c_{p_{1}+1}+c_{j}))^{2}}{(\sigma^{2}_{p_{1}+1}/2c_{p_{1}+1})(\sigma^{2}_{j}/2c_{j})}\right)
	\label{eq:eq70}
\end{align}
\noindent 
Comparing (\ref{eq:eq67}) with (\ref{eq:eq68}) and (\ref{eq:eq69}) with (\ref{eq:eq70}) 
implies that $\hat{j}_{n}$ will asymptotically point to either $j=1$ or $j=p_{1}+1$ (i.e. one of the two true predictors) as stated. \hfill $\blacksquare$

\section{Further Experimental Properties}

This section provides additional simulation based results extending the size/power based outcomes presented in Section 7 of the main paper. 

\subsection{Empirical Size}

The supplementary size based simulations use the same DGP as in Section 7 (Table 1) and aim to illustrate the influence of alternative variance normalisers on size (i.e., using the formulation of $\hat{\omega}_{n}^{2,a}$ in (26) that is based on residuals under the null hypothesis instead of $\hat{\omega}_{n}^{2,b}$ in (27) that was used in Table 1 of the main paper).

To highlight the role played by the size-neutral power enhancing transformation introduced in Section 5 we also present corresponding empirical size outcomes based on the unadjusted ${\cal D}_{n}(\mu_{0})$ statistic. For notational purposes we refer to the power enhanced/size-neutral test statistic 
evaluated using the variance normalisers $\hat{\omega}_{n}^{2,a}$ and $\hat{\omega}_{n,j}^{2,b}$ as ${\cal D}_{n}^{d,a}(\mu_{0})$
and ${\cal D}_{n}^{d,b}(\mu_{0})$ respectively. Their non-enhanced versions are denoted  
${\cal D}_{n}^{a}(\mu_{0})$
and ${\cal D}_{n}^{b}(\mu_{0})$. All our simulations in the main text have been obtained using 
${\cal D}_{n}^{d,b}(\mu_{0})$ so that in what follows we compare outcomes with  
${\cal D}_{n}^{d,a}(\mu_{0})$ (adjusted, null residuals), ${\cal D}_{n}^{a}(\mu_{0})$ (unadjusted, null residuals) and ${\cal D}_{n}^{b}(\mu_{0})$ (unadjusted, residuals under alternative). 

Table \ref{tab:TabB1} presents empirical size outcomes based on the power enhanced statistic as in the main text  but using a variance normaliser based on the null residuals (i.e., $\hat{\omega}^{2}_{n}=\hat{\omega}^{2,a}_{n}$) and can be compared with Table 1 in the main text which was based on  $\hat{\omega}^{2}_{n}=\hat{\omega}^{2,b}_{n,j}$. We note virtually identical outcomes across all DGP scenarios suggesting that when it comes to the size of our proposed test statistic the use of either 
$\hat{\omega}^{2,a}_{n}$ or $\hat{\omega}^{2,b}_{n,j}$ makes little practical difference. All empirical size outcomes match their nominal counterparts very closely and often nearly perfectly regardless of which variance normaliser is used.  

Tables \ref{tab:TabB2} and \ref{tab:TabB3} have repeated the same exercise using the ``raw'' (unadjusted) version of our test statistic. Although here we continue to note very little difference in outcomes based on either 
$\hat{\omega}^{2,a}_{n}$ or $\hat{\omega}^{2,b}_{n,j}$ the main message that comes across these two tables is the importance and effectiveness of our proposed adjustment. The size properties of the 
unadjusted statistics clearly deteriorate as $\mu_{0}\rightarrow 0.5$ (i.e., the variance degeneracy bound) despite remaining unaffected by the number of predictors under consideration. Under $p=500$ for instance empirical size is in the vicinity of 8\% for $\mu_{0}=0.35$ but drops to about 3\% for $\mu_{0}=0.45$. Our adjusted statistic is not subject to such distortions and shows a remarkably effective ability to align itself with its nominal counterparts regardless of DGP parameterisations. 
Although our proposed adjustment is solely designed to enhance power it is also clear that it 
helps maintain good finite sample size, a feature motivated in Remark 4 in the main text.  

\setcounter{table}{0}
\renewcommand{\thetable}{B\arabic{table}}

\begin{table}[h]
	\centering
	\caption{Empirical Size of ${\cal D}_{n}^{d,a}(\mu_{0})$ (10\% Nominal)} \vspace{0.5cm}
	\scalebox{0.88}{	\begin{tabular}{lccccccccc} 
			$\mu_{0}$ & p=10  & p=50  & p=500 & p=10  & p=50  & p=500 & p=10  & p=50  & p=500  \\ \hline 
			& \multicolumn{3}{c}{A(i)} & \multicolumn{3}{c}{A(ii)} & \multicolumn{3}{c}{A(iii)} \\
			0.35  & 0.106 & 0.103 & 0.103 & 0.107 & 0.104 & 0.104 & 0.106 & 0.105 & 0.104 \\
			0.40  & 0.108 & 0.104 & 0.104 & 0.109 & 0.106 & 0.106 & 0.107 & 0.106 & 0.106 \\
			0.45  & 0.110 & 0.093 & 0.100 & 0.115 & 0.094 & 0.101 & 0.115 & 0.094 & 0.101 \\
			& \multicolumn{3}{c}{B(i)} & \multicolumn{3}{c}{B(ii)} & \multicolumn{3}{c}{B(iii)} \\
			0.35  & 0.106 & 0.104 & 0.103 & 0.107 & 0.103 & 0.104 & 0.109 & 0.104 & 0.104 \\
			0.40  & 0.115 & 0.106 & 0.105 & 0.115 & 0.105 & 0.105 & 0.115 & 0.105 & 0.105 \\
			0.45  & 0.116 & 0.099 & 0.102 & 0.119 & 0.100 & 0.104 & 0.122 & 0.100 & 0.104 \\
			& \multicolumn{3}{c}{C(i)} & \multicolumn{3}{c}{C(ii)} & \multicolumn{3}{c}{C(iii)} \\
			0.35  & 0.107 & 0.104 & 0.103 & 0.106 & 0.103 & 0.103 & 0.107 & 0.103 & 0.103 \\
			0.40  & 0.111 & 0.105 & 0.104 & 0.110 & 0.107 & 0.104 & 0.110 & 0.107 & 0.104 \\
			0.45  & 0.116 & 0.093 & 0.101 & 0.118 & 0.098 & 0.102 & 0.119 & 0.099 & 0.102 \\
	\end{tabular}}
	\label{tab:TabB1}
\end{table}

\begin{table}[h]
	\centering
	\caption{Empirical Size of ${\cal D}_{n}^{a}(\mu_{0})$ (10\% Nominal)} \vspace{0.5cm}
	\scalebox{0.88}{	\begin{tabular}{lccccccccc}
			& \multicolumn{3}{c}{A(i)} & \multicolumn{3}{c}{A(ii)} & \multicolumn{3}{c}{A(iii)} \\ \hline
			$\mu_{0}$ & p=10  & p=50  & p=500 & p=10  & p=50  & p=500 & p=10  & p=50  & p=500 \\
			0.35  & 0.080 & 0.078 & 0.076 & 0.080 & 0.078 & 0.076 & 0.080 & 0.078 & 0.076 \\
			0.40  & 0.069 & 0.063 & 0.065 & 0.067 & 0.062 & 0.065 & 0.068 & 0.062 & 0.065 \\
			0.45  & 0.035 & 0.031 & 0.031 & 0.036 & 0.032 & 0.031 & 0.036 & 0.032 & 0.031 \\
			& \multicolumn{3}{c}{B(i)} & \multicolumn{3}{c}{B(ii)} & \multicolumn{3}{c}{B(iii)} \\
			0.35  & 0.074 & 0.073 & 0.069 & 0.074 & 0.073 & 0.070 & 0.073 & 0.073 & 0.070 \\
			0.40  & 0.061 & 0.054 & 0.059 & 0.059 & 0.054 & 0.059 & 0.060 & 0.053 & 0.059 \\
			0.45  & 0.026 & 0.023 & 0.024 & 0.030 & 0.024 & 0.024 & 0.030 & 0.024 & 0.024 \\
			& \multicolumn{3}{c}{C(i)} & \multicolumn{3}{c}{C(ii)} & \multicolumn{3}{c}{C(iii)} \\
			0.35  & 0.078 & 0.075 & 0.073 & 0.077 & 0.076 & 0.073 & 0.077 & 0.076 & 0.073 \\
			0.40  & 0.063 & 0.058 & 0.061 & 0.065 & 0.059 & 0.061 & 0.065 & 0.059 & 0.061 \\
			0.45  & 0.030 & 0.027 & 0.027 & 0.032 & 0.027 & 0.026 & 0.031 & 0.027 & 0.026 \\
	\end{tabular}}
	\label{tab:TabB2}
\end{table}

\begin{table}[h]
	\centering
	\caption{Empirical Size of ${\cal D}_{n}^{b}(\mu_{0})$ (10\% Nominal)} \vspace{0.5cm}
	\scalebox{0.88}{	\begin{tabular}{lccccccccc}
			$\mu_{0}$ & p=10  & p=50  & p=500 & p=10  & p=50  & p=500 & p=10  & p=50  & p=500 \\ \hline
			& \multicolumn{3}{c}{A(i)} & \multicolumn{3}{c}{A(ii)} & \multicolumn{3}{c}{A(iii)} \\
			0.35  & 0.080 & 0.077 & 0.075 & 0.080 & 0.078 & 0.076 & 0.079 & 0.078 & 0.076 \\
			0.40  & 0.069 & 0.062 & 0.064 & 0.067 & 0.062 & 0.065 & 0.068 & 0.062 & 0.065 \\
			0.45  & 0.035 & 0.031 & 0.031 & 0.036 & 0.032 & 0.031 & 0.037 & 0.032 & 0.031 \\
			& \multicolumn{3}{c}{B(i)} & \multicolumn{3}{c}{B(ii)} & \multicolumn{3}{c}{B(iii)} \\
			0.35  & 0.074 & 0.072 & 0.068 & 0.074 & 0.072 & 0.069 & 0.073 & 0.072 & 0.069 \\
			0.40  & 0.060 & 0.054 & 0.058 & 0.059 & 0.054 & 0.059 & 0.059 & 0.053 & 0.059 \\
			0.45  & 0.026 & 0.023 & 0.024 & 0.030 & 0.024 & 0.025 & 0.030 & 0.024 & 0.024 \\
			& \multicolumn{3}{c}{C(i)} & \multicolumn{3}{c}{C(ii)} & \multicolumn{3}{c}{C(iii)} \\
			0.35  & 0.077 & 0.075 & 0.071 & 0.076 & 0.076 & 0.072 & 0.076 & 0.076 & 0.072 \\
			0.40  & 0.062 & 0.057 & 0.061 & 0.065 & 0.059 & 0.061 & 0.065 & 0.058 & 0.061 \\
			0.45  & 0.030 & 0.027 & 0.027 & 0.032 & 0.027 & 0.026 & 0.031 & 0.027 & 0.026 \\
	\end{tabular}}
	\label{tab:TabB3}
\end{table}

\subsection{Empirical Power}

Our supplementary power based simulations follow the same logical flow as above. They aim to illustrate the role of using an alternative variance normaliser, namely $\hat{\omega}^{2,a}_{n}$ instead of $\hat{\omega}^{2,b}_{n}$. As the former is based on the use of residuals under the null 
we expect potentially important finite sample differences in power performance between 
${\cal D}_{n}^{d,b}(\mu_{0})$ used in the main text and ${\cal D}_{n}^{d,a}(\mu_{0})$ considered here. The simulations that follow also illustrate the important role played by our power-enhancing 
adjustment by repeating all experiments across adjusted and unadjusted versions of our test statistic. Although the outcomes presented below are based on the same DGPs as in the main text (numbered as (i), (ii-a) and (ii-b)) we also include an additional scenario (referred to as (iii) in what follows) that mixes both stationary and persistent predictors within (30). For this purpose we let 
$\beta_{a}^{*} \in \{(2,3,4,5)\}$, $\beta_{b}^{*} \in \{(5,6,7,8)\}$, $\beta_{c}^{*} \in \{(2,3,4,5)\}$, $\beta_{d}^{*} \in \{(5,6,7,8)\}$ so that the DGP as defined in (30) includes a total of four predictors, two of which are purely stationary and the remaining two persistent. 

\begin{table}[h]
	\centering
	\caption{Empirical Power of ${\cal D}_{n}^{d,a}(\mu_{0})$ under DGPs (i)-(ii)}
	\begin{tabular}{lcccc} \hline
		& \multicolumn{4}{c}{DGP (i)} \\
		$\beta_{an}$ & 0.423 & 0.634 & 0.846 & 1.057 \\
		$\beta_{bn}$ & 1.057 & 1.269 & 1.480 & 1.692 \\ \hline 
		$\mu_{0}=0.35$ & 0.487 & 0.530 & 0.572 & 0.589 \\
		$\mu_{0}=0.40$ & 0.694 & 0.753 & 0.793 & 0.819 \\
		$\mu_{0}=0.45$ & 0.977 & 0.992 & 0.996 & 0.998 \\ \hline \hline
		& \multicolumn{4}{c}{DGP (ii-a)} \\
		$\beta_{cn}$ & 0.030 & 0.045 & 0.060 & 0.075 \\
		$\beta_{dn}$ & 0.075 & 0.090 & 0.106 & 0.121 \\ \hline
		$\mu_{0}=0.35$ & 0.171 & 0.224 & 0.292 & 0.329 \\
		$\mu_{0}=0.40$ & 0.211 & 0.286 & 0.368 & 0.434 \\
		$\mu_{0}=0.45$ & 0.337 & 0.479 & 0.611 & 0.707 \\ \hline \hline
		& \multicolumn{4}{c}{DGP (ii-b)} \\
		$\beta_{cn}$ & 0.075 & 0.090 & 0.106 & 0.121 \\
		$\beta_{dn}$ & 0.121 & 0.136 & 0.151 & 0.166 \\ \hline
		$\mu_{0}=0.35$ & 0.317 & 0.372 & 0.427 & 0.451 \\
		$\mu_{0}=0.40$ & 0.427 & 0.491 & 0.554 & 0.590 \\
		$\mu_{0}=0.45$ & 0.704 & 0.779 & 0.839 & 0.870 
	\end{tabular}
	\label{tab:TabB4}
\end{table}

Comparing Tables \ref{tab:TabB4}-\ref{tab:TabB5} with Table 2 in the main text highlights the unfavourable influence on power 
of using a variance normaliser based on the null residuals. Take for instance the 
case of DGP (i) with $\mu_{0}=0.40$ for which we have an empirical power estimate of 69.4\% in 
Table \ref{tab:TabB4} based on ${\cal D}_{n}^{d,a}(\mu_{0})$. This can be compared with 
the estimate of 97.7\% obtained using ${\cal D}_{n}^{d,b}(\mu_{0})$ (Table 2 of main text). 
The fact that power is much superior when using $\hat{\omega}^{2,b}_{nj}$ as the variance normaliser and the DGP is driven by purely stationary predictors is a well known phenomenon that has been widely documented in contexts such as change-point detection with cusum type statistics. The mere fact that the variance normaliser takes into account information under the alternative acts as an important power booster. 

Interestingly, these power differences arise solely in the context of DGP(i) that is driven solely by purely stationary predictors. If we compare Table \ref{tab:TabB4} with Table 2 for DGPs (ii-a) and (ii-b) which contain solely persistent predictors we note very similar power estimates regardless of whether inferences are based on  ${\cal D}_{n}^{d,a}(\mu_{0})$ or  ${\cal D}_{n}^{d,b}(\mu_{0})$. 

Tables \ref{tab:TabB5}-\ref{tab:TabB6} repeat the above power experiments using unadjusted test statistics. Table \ref{tab:TabB5} is based on ${\cal D}_{n}^{a}(\mu_{0})$ that uses null residuals 
while Table \ref{tab:TabB6} is based on ${\cal D}_{n}^{b}(\mu_{0})$ that uses residuals from the larger model. Outcomes strongly support the use of our power enhanced formulation. Comparing Table  
\ref{tab:TabB6} with Table 2 in the main text for instance and focusing on their last columns 
we note that the power enhancement boosts empirical powers by more than 20\%. A similar picture can also be observed from \ref{tab:TabB7} based on DGP(iii) described above. Comparing outcomes on its mid-panel based on ${\cal D}_{n}^{b}(\mu_{0})$ (unadjusted) with its bottom panel based on 
${\cal D}_{n}^{d,b}(\mu_{0})$ (adjusted) we note substantial spreads for low signal to noise 
parameterisations. As the signal to noise ratio increases these differences tend to progressively  dissipate but remain non-negligible (e.g., 83.5\% versus 99.9\% under the most favourable signal to noise scenario).

\begin{table}[h]
	\centering
	\caption{Empirical Power of ${\cal D}_{n}^{a}(\mu_{0})$ under DGPs (i)-(ii)}
	\begin{tabular}{lcccc} \hline \hline 
		& \multicolumn{4}{c}{DGP (i)} \\
		$\beta_{an}$ & 0.423 & 0.634 & 0.846 & 1.057 \\
		$\beta_{bn}$ & 1.057 & 1.269 & 1.480 & 1.692 \\ \hline 
		$\mu_{0}=0.35$ & 0.224 & 0.241 & 0.258 & 0.258 \\
		$\mu_{0}=0.40$ & 0.276 & 0.299 & 0.333 & 0.337 \\
		$\mu_{0}=0.45$ & 0.451 & 0.520 & 0.577 & 0.592 \\ \hline \hline
		& \multicolumn{4}{c}{DGP (ii-a)} \\
		$\beta_{cn}$ & 0.030 & 0.045 & 0.060 & 0.075 \\
		$\beta_{dn}$ & 0.075 & 0.090 & 0.106 & 0.121 \\ \hline 
		$\mu_{0}=0.35$ & 0.093 & 0.124 & 0.161 & 0.177 \\
		$\mu_{0}=0.40$ & 0.089 & 0.125 & 0.164 & 0.192 \\
		$\mu_{0}=0.45$ & 0.070 & 0.118 & 0.179 & 0.231 \\ \hline \hline
		& \multicolumn{4}{c}{DGP (ii-b)} \\
		$\beta_{cn}$ & 0.075 & 0.090 & 0.106 & 0.121 \\
		$\beta_{dn}$ & 0.121 & 0.136 & 0.151 & 0.166 \\ \hline 
		$\mu_{0}=0.35$ & 0.169 & 0.209 & 0.243 & 0.261 \\
		$\mu_{0}=0.40$ & 0.180 & 0.225 & 0.265 & 0.295 \\
		$\mu_{0}=0.45$ & 0.221 & 0.281 & 0.336 & 0.377 \\
	\end{tabular}
	\label{tab:TabB5}
\end{table}

\begin{table}[h]
	\centering
	\caption{Empirical Power of ${\cal D}_{n}^{b}(\mu_{0})$ under DGPs (i)-(ii)}
	\begin{tabular}{lcccc} \hline 
		& \multicolumn{4}{c}{DGP (i)} \\
		$\beta_{an}$ & 0.423 & 0.634 & 0.846 & 1.057 \\
		$\beta_{bn}$ & 1.057 & 1.269 & 1.480 & 1.692 \\ \hline
		$\mu_{0}=0.35$ & 0.407 & 0.519 & 0.613 & 0.679 \\
		$\mu_{0}=0.40$ & 0.577 & 0.739 & 0.834 & 0.896 \\
		$\mu_{0}=0.45$ & 0.918 & 0.983 & 0.997 & 1.000 \\ \hline \hline
		& \multicolumn{4}{c}{DGP (ii-a)} \\
		$\beta_{cn}$ & 0.030 & 0.045 & 0.060 & 0.075 \\
		$\beta_{dn}$ & 0.075 & 0.090 & 0.106 & 0.121 \\ \hline 
		$\mu_{0}=0.35$ & 0.093 & 0.128 & 0.167 & 0.187 \\
		$\mu_{0}=0.40$ & 0.091 & 0.128 & 0.172 & 0.204 \\
		$\mu_{0}=0.45$ & 0.073 & 0.128 & 0.198 & 0.262 \\ \hline \hline
		& \multicolumn{4}{c}{DGP (ii-b)} \\
		$\beta_{cn}$ & 0.075 & 0.090 & 0.106 & 0.121 \\
		$\beta_{dn}$ & 0.121 & 0.136 & 0.151 & 0.166 \\ \hline 
		$\mu_{0}=0.35$ & 0.178 & 0.221 & 0.261 & 0.286 \\
		$\mu_{0}=0.40$ & 0.195 & 0.241 & 0.291 & 0.323 \\
		$\mu_{0}=0.45$ & 0.252 & 0.323 & 0.395 & 0.445 \\
	\end{tabular}
	\label{tab:TabB6}
\end{table}

\begin{table}[htbp]
	\centering
	\caption{Empirical Power under DGP(iii)}
	\scalebox{1}{\begin{tabular}{lcccc} \hline
			$\beta_{an}$ & 0.423 & 0.634 & 0.846 & 1.057 \\
			$\beta_{bn}$ & 1.057 & 1.269 & 1.480 & 1.692 \\
			$\beta_{cn}$ & 0.030 & 0.045 & 0.060 & 0.075 \\
			$\beta_{dn}$ & 0.075 & 0.090 & 0.106 & 0.121 \\ \hline \hline
			& \multicolumn{4}{c}{${\cal D}_{n}^{a}(\mu_{0})$} \\
			$\mu_{0}=0.35$ & 0.219 & 0.245 & 0.255 & 0.270 \\
			$\mu_{0}=0.40$ & 0.270 & 0.307 & 0.330 & 0.355 \\
			$\mu_{0}=0.45$ & 0.464 & 0.542 & 0.575 & 0.607 \\ \hline
			& \multicolumn{4}{c}{${\cal D}_{n}^{b}(\mu_{0})$} \\
			$\mu_{0}=0.35$ & 0.382 & 0.488 & 0.559 & 0.605 \\
			$\mu_{0}=0.40$ & 0.554 & 0.695 & 0.789 & 0.835 \\
			$\mu_{0}=0.45$ & 0.905 & 0.974 & 0.994 & 0.997 \\ \hline
			& \multicolumn{4}{c}{${\cal D}_{n}^{d,a}(\mu_{0})$} \\
			$\mu_{0}=0.35$ & 0.489 & 0.544 & 0.566 & 0.583 \\
			$\mu_{0}=0.40$ & 0.710 & 0.768 & 0.795 & 0.823 \\
			$\mu_{0}=0.45$ & 0.982 & 0.992 & 0.996 & 0.998 \\ \hline
			& \multicolumn{4}{c}{${\cal D}_{n}^{d,b}(\mu_{0})$} \\
			$\mu_{0}=0.35$ & 0.807 & 0.913 & 0.957 & 0.975 \\
			$\mu_{0}=0.40$ & 0.972 & 0.995 & 0.999 & 0.999 \\
			$\mu_{0}=0.45$ & 1.000 & 1.000 & 1.000 & 1.000 \\
	\end{tabular}}
	\label{tab:TabB7}
\end{table}

\end{document}